\documentclass[fleqn,usenatbib]{mnras}

\usepackage{newtxtext,newtxmath}
\usepackage[T1]{fontenc}
\usepackage{physics}

\DeclareRobustCommand{\VAN}[3]{#2}
\let\VANthebibliography\thebibliography
\def\thebibliography{\DeclareRobustCommand{\VAN}[3]{##3}\VANthebibliography}

\usepackage{graphicx}	% Including figure files
\usepackage{amsmath}	% Advanced maths commands
\title[Hydrogen-silane-water envelopes of sub-Neptunes]{Atmospheres as windows into sub-Neptune interiors: coupled chemistry and structure of hydrogen-silane-water envelopes}

\author[W. Misener, H.E. Schlichting and E.D. Young]{
William Misener$^{1}$, Hilke E. Schlichting$^{1}$ and Edward D. Young$^{1}$
\\
$^{1}$Department of Earth, Planetary, and Space Sciences, The University of California, Los Angeles, 595 Charles E. Young Drive East, Los Angeles, CA 90095, USA\\}

\date{Accepted XXX. Received YYY; in original form ZZZ}

\pubyear{2023}

\begin{document}
\label{firstpage}
\pagerange{\pageref{firstpage}--\pageref{lastpage}}
\maketitle

% Abstract of the paper
\begin{abstract}
Sub-Neptune exoplanets are commonly hypothesized to consist of a silicate-rich magma ocean topped by a hydrogen-rich atmosphere. Previous work studying the outgassing of silicate material has demonstrated that such atmosphere-interior interactions can affect the atmosphere's overall structure and extent. But these models only considered SiO in an atmosphere of hydrogen gas, without considering chemical reactions between them. Here we couple calculations of the chemical equilibrium between H, Si, and O species with an atmospheric structure model. We find that substantial amounts of silane, SiH$_4$, and water, H$_2$O, are produced by the interaction between the silicate-rich interior and hydrogen-rich atmosphere. These species extend high into the atmosphere, though their abundance is greatest at the hottest, deepest regions. For example, for a 4~$M_\oplus$ planet with an equilibrium temperature of 1000~K, a base temperature of 5000~K, and a 0.1~$M_\oplus$ hydrogen envelope, silicon species and water can comprise 30 percent of the atmosphere by number at the bottom of the atmosphere. Due to this abundance enhancement, we find that convection is inhibited at temperatures $\gtrsim 2500$~K. This temperature is lower, implying that the resultant non-convective region is thicker, than was found in previous models which did not account for atmospheric chemistry. Our findings show that significant endogenous water is produced by magma-hydrogen interactions alone, without the need to accrete ice-rich material. We discuss the observability of the signatures of atmosphere-interior interaction and directions for future work, including condensate lofting and more complex chemical networks.
\end{abstract} 

% Select between one and six entries from the list of approved keywords.
% Don't make up new ones.

%%%%%%%%%%%%%%%%%%%%%%%%%%%%%%%%%%%%%%%%%%%%%%%%%%

%%%%%%%%%%%%%%%%% BODY OF PAPER %%%%%%%%%%%%%%%%%%

\section{Introduction}
Exoplanet surveys have revealed that planets with radii between 1 and 4 Earth radii with orbital periods shorter than 100 days are the most intrinsically common type of planet yet observed \citep[e.g.][]{Fressin13}. Precise mass and radius measurements indicate that these planets are distributed bimodally in radius and bulk density \citep{WM14,Fulton17}. These measurements separate the small planet population into super-Earths, smaller planets consistent with bulk-Earth composition, and larger sub-Neptunes, which must contain some low-density material to explain their measured radii and densities.

Typically, sub-Neptunes are assumed to contain some combination of terrestrial rock and metal, icy material, and/or hydrogen gas \citep[e.g.][]{RS10,D17,Zeng2019}. As hydrogen is the lowest density of these three materials, an envelope containing a small amount of it, of order one percent of a planet's total mass, can greatly increase a planet's size and thus decrease its bulk density. It is therefore possible to model most sub-Neptunes as Earth-like cores with hydrogen envelopes of a few percent the planet's mass \citep[e.g.][]{LF14}. However, density measurements do not rule out compositions with large mass fractions in ices, such as water, with correspondingly smaller hydrogen atmospheres \citep[e.g.][]{Zeng2019,LuquePalle2022}.

Due to their ubiquity, a large number of models of sub-Neptune atmospheric evolution and composition have been put forward. Crucially, neither the radii nor masses of these envelopes are expected to remain constant in time. Rather, sub-Neptune atmospheres shrink in extent as the planets cool into space \citep{LF14}, and they can be susceptible to atmospheric stripping. This stripping, which can be due to either photo-evaporative \citep{OJ12,OW17} or core-powered mass-loss mechanisms \citep{GSS16,GS19}, is thought to have stripped some sub-Neptunes entirely, turning them into super-Earths and forming the observed radius valley. The sub-Neptunes that remain were able to mostly resist this stripping. The attributes of the observed radius valley are best matched by mass-loss models if the cores of sub-Neptunes are mostly rocky, with little ice \citep{GS19,RO21}.

However, these and other models typically assume each compositional constituent of the planet is contained its own layer. Such a structure may be the simplest model to first order, but it may not accurately describe the interactions between these constituents at the high temperatures and pressures expected within sub-Neptunes. Awareness of mixing between layers previously modeled as separate is gaining traction across planetary science. In the Solar System, Jupiter and Saturn show evidence for non-discrete cores which have blended with their metallic H surroundings \citep{W17Juno,MankovichFuller21Saturn}. Similar mixing between water and hydrogen has been proposed in the ice giants \citep{BaileyStevenson2021}, and water and rock may be miscible at the temperature-pressure conditions of sub-Neptune interiors \citep{VazanSari22}.

Another blurred line between layers arises from the interaction between a hydrogen atmosphere and the potential rocky, silicate-dominated core beneath it. Recent work has shown that at chemical equilibrium, significant silicate vapor is stable in the gas at the base of young sub-Neptune atmospheres \citep{SY22}, where temperatures can exceed 5000~K \citep{GSS16}. As silicate vapor will decline in abundance with decreasing temperatures, this implies a compositional gradient deep within sub-Neptunes. This gradient in composition, and thus in mean molecular weight, has been demonstrated to inhibit convection \citep{MS22,Markham22}. The inhibition occurs because a deeper parcel is heavier than one higher up, which overcomes its thermal buoyancy. Such an effect has long been known to apply at the conditions within the Solar system gas and ice giants, where the condensable considered is usually water vapor \citep{G95,L17,Markham21}. But it has only recently been applied to sub-Neptune planets with magma oceans. In addition to arguments from chemical equilibrium, gradients in Si abundance in a hydrogen dominated atmosphere, and therefore a non-convective region, could also form as a consequence of the accretion of pebbles during formation \citep{BrouwersOrmel2020,OrmelVazan2021}, though the subsequent evolution of such structures remain unclear.

These studies mark initial forays into understanding the interiors of this ubiquitous class of planet, sub-Neptunes composed of hydrogen and silicate. Much work remains to be done to further quantify the effects of interaction between the interior and atmosphere. In particular, both \citet{MS22} and \citet{Markham22} assume the gas released into the atmosphere is pure SiO vapor. However, oxidized SiO is not inert in a hydrogen-dominated background gas. Rather, it will react with the hydrogen, producing water (H$_2$O) and silane (SiH$_4$). These species will alter the impact of magma condensation on the overall atmospheric structure. It will also change the observable signatures of interior-atmosphere interactions we expect. In this paper, we construct a coupled atmospheric-chemical model which captures the interactions we expect vaporizing silicate magma to have with a hydrogen atmosphere.

\section{Model}
In this section we detail our atmospheric and chemical model of a sub-Neptune consisting of a silicate-dominated interior and a hydrogen-dominated atmosphere.

\subsection{Chemistry}\label{sec:chem}
In \citet{MS22}, to quantify the partial pressure of rock vapor in the atmosphere above a pure SiO$_2$ magma ocean, the authors use an exponential relationship between SiO and temperature appropriate for congruent evaporation of SiO$_2$ melt, taken from \citet{VF13}. The evaporation reaction can be written as:
\begin{equation}\label{eq:SiOrxn}\tag{R1}
    \mathrm{SiO}_{2,l} \rightleftharpoons \mathrm{SiO} + 0.5 \mathrm{O}_{2},
\end{equation}
where on the left hand side, the $l$ subscript denotes a liquid species; here and going forward, all chemical species without this subscript are gaseous. Reaction~\ref{eq:SiOrxn} can be quantified using an equilibrium constant:
\begin{equation}\label{eq:SiOK}
    K_{\mathrm{eq, R1}} = \frac{P_\mathrm{SiO} P_\mathrm{O_2}^{0.5}}{a_\mathrm{SiO_2}},
\end{equation}
where $P_i$ denotes the partial pressure of species $i$, and $a_\mathrm{SiO_2}$ is the activity of SiO$_2$ in the melt, which we take to be 1, i.e., we assume the silicate melt is fully SiO$_2$. This and following equilibrium constants are found by taking the difference between the Gibbs free energies of the products and reactants. All Gibbs free energy values are calculated from the NIST Chemistry WebBook parameterizations of enthalpy and entropy, based on data from the JANAF tables \citep{NIST}. For SiO$_{2, l}$, the NIST data extends to a maximum temperature of 4500~K; in this work we extrapolate the NIST fit to 5000~K, following \citet{SY22}. We also extrapolate the NIST fit below 1996~K, where solid SiO$_{2, s}$ should be the predominant species. We verify that using the NIST values for the Gibbs free energy of the solid form does not alter our results. The equilibrium constant is a strong function of temperature, where higher temperatures produce relatively more evaporation and thus higher partial pressures of the gaseous species. We demonstrate the temperature dependence of $K_\mathrm{eq, R1}$, as well as the equilibrium constants defined in Eqs.~\ref{eq:O2K} and~\ref{eq:SiH4K} below, in Appendix~\ref{sec:Keq_plots}. Application of this equilibrium constant produces good agreement with the \citet{VF13} equation, though there are slight differences due to the different thermodynamic values used.

However, both species on the right-hand side of Reaction~\ref{eq:SiOrxn} are oxidized, and therefore neither is expected to be inert in the presence of H$_2$ gas, a strong reducer. Rather, each product should interact with the background hydrogen. Oxygen combines with hydrogen to produce water
\begin{equation}\label{eq:O2rxn}\tag{R2}
    0.5 \mathrm{O}_2 + \mathrm{H}_{2} \rightleftharpoons \mathrm{H}_2\mathrm{O}
\end{equation}
while SiO reacts with hydrogen to produce silane, SiH$_4$
\begin{equation}\label{eq:SiH4rxn}\tag{R3}
    \mathrm{SiO} + 3 \mathrm{H}_2 \rightleftharpoons \mathrm{SiH}_4 + \mathrm{H}_2\mathrm{O}.
\end{equation}

The production of silane  has been suggested to occur deep within Jupiter, at low concentrations \citep[e.g.][]{FegleyLodders1994}, due to silane being more thermochemically favorable than SiO at the relevant temperature-pressure conditions \citep{Visscher2010}. More recently, the same reaction has been suggested to occur in sub-Neptune atmospheres which interface with magma oceans \citep{Markham22}, where plentiful oxidized silicate is presumed to exist. Moreover, diamond-anvil cell experiments indicate that mixtures of SiO$_2$ and H$_2$ fluid produce silane and water in the fluid at $2 \times 10^4$ bar and $1700$ K \citep{Shinozaki14}, and similar silane production was observed in a MgSiO$_3$--H$_2$ mixture at $3.6 \times 10^4$ bar and $2000$ K \citep{Shinozaki16}. These pressures and temperatures are relevant both to the inner mantle of Earth and to the sub-Neptune conditions we consider here (see Fig.~\ref{fig:prof} below).

The relative abundances of these species in the atmosphere are determined by the equilibrium constants for the reactions, respectively:
\begin{equation}\label{eq:O2K}
    K_{\mathrm{eq, R2}} = \frac{P_\mathrm{H_2O}} {P_\mathrm{H_2} P_\mathrm{O_2}^{0.5}}
\end{equation}
and 
\begin{equation}\label{eq:SiH4K}
    K_{\mathrm{eq, R3}} = \frac{P_\mathrm{SiH_4} P_\mathrm{H_2O}}{P_\mathrm{SiO} P_\mathrm{H_2}^3}.
\end{equation}
As with Eq.~\ref{eq:SiOK}, these equilibrium constants are calculated from Gibbs free energies taken from NIST and are functions of temperature.

To calculate the atmospheric composition at each level of the atmosphere, we use the temperature, $T$, and total pressure, $P$, derived from the atmospheric structure equations detailed below in Sec.~\ref{sec:structure}. We then apply the fact that in our model, the changes in atmospheric composition are driven by the evaporation and condensation of SiO$_2$. Therefore, we can use the stoichiometric relationship that for every Si atom at a given level in the atmosphere, there must be two O atoms. More quantitatively, we can calculate the ``partial pressures'' of each element, which both convert through common factors to the numbers at each pressure level, 
\begin{equation}\label{eq:sumPSi}
    \sum P_\mathrm{Si} \equiv P_\mathrm{SiO} + P_\mathrm{SiH_4}
\end{equation}
and 
\begin{equation}\label{eq:sumPO}
    \sum P_\mathrm{O} \equiv P_\mathrm{SiO} + 2 P_\mathrm{O_2} + P_\mathrm{H_2O}.
\end{equation}
We can then mandate that 
\begin{equation}\label{eq:Nrelation}
   \sum P_\mathrm{O} = 2 \sum P_\mathrm{Si}
\end{equation}
at each level of the atmosphere.

Eqs.~\ref{eq:SiOK}, \ref{eq:O2K}, \ref{eq:SiH4K}, and \ref{eq:Nrelation} are sufficient to solve for the partial pressures of each component. We detail our analytic method in Appendix~\ref{sec:partialpress}.

\subsection{Atmospheric Structure}\label{sec:structure}
This atmospheric structure model builds on \citet{MS22}, which added consideration of compositional gradients and moist convection to the model of \citet{MS21}. Here, as in both of those works, we model the outer atmosphere as being in thermal equilibrium with the incident flux from the star and thus isothermal at the planet's equilibrium temperature, $T_\mathrm{eq}$ \citep[e.g.][]{LC15,GSS16}.

The atmosphere transitions to convective at the outer radiative-convective boundary ($R_\mathrm{rcb}$). The radial pressure gradient is found following hydrostatic equilibrium:
\begin{equation}\label{eq:dPdr}
    \frac{\partial P}{\partial R} = -\frac{G M_\mathrm{c}}{R^2} \frac{\mu P}{k_\mathrm{B} T},
\end{equation}
where $G$ is the gravitational constant, $M_\mathrm{c}$ is the planet mass, $k_\mathrm{B}$ is the Boltzmann constant, and $\mu$ is the local mean molecular weight. The mean molecular weight is a function of temperature and pressure, and is calculated using chemical equilibrium, as detailed in the previous section.

The temperature profile in the convective region follows a moist adiabat:
\begin{equation}\label{eq:wetadiabat}
    \frac{\partial \ln T}{\partial \ln P} = \frac{k_\mathrm{B}}{\mu} \cfrac{1+\cfrac{P_\mathrm{Si}}{P_\mathrm{H}} \cfrac{\partial \ln P_\mathrm{Si}} {\partial T}} {c_\mathrm{p} +\cfrac{P_\mathrm{Si}} {P_\mathrm{H}} \cfrac{k_\mathrm{B}}{\mu} T^2 \bigg(\cfrac{\partial \ln P_\mathrm{Si}} {\partial T}\bigg)^2},
\end{equation}
where $P_\mathrm{H}$ is the partial pressure of hydrogen, and $P_\mathrm{Si}$ is the combined partial pressures of Si-bearing species \citep[e.g.][]{L17}. The determination of these partial pressures via chemical equilibrium was described in Section \ref{sec:chem}. The specific heat capacity $c_\mathrm{p}$ is a mass-weighted average of the specific heat capacities of each component:
\begin{equation}
    c_\mathrm{p} = \sum_i q_i c_{\mathrm{p}, i}
\end{equation}
where $q_i$ is the mass mixing ratio, defined as $q_i \equiv \mu_i P_i/(\mu P)$, with $\mu_i$ being the molecular weight of species $i$. $c_{\mathrm{p}, i}$, the heat capacity of each species, is given by
\begin{equation}
    c_{\mathrm{p}, i} = \frac{k_\mathrm{B}}{\mu_i}\frac{\gamma_i}{\gamma_i-1}.
\end{equation}
where $\gamma_i$ is the adiabatic index of a species. For the diatomic molecules O$_2$, H$_2$, and SiO, we assume $\gamma=7/5$, while for H$_2$O we take $\gamma=4/3$ and for SiH$_4$ we take $\gamma=1.3$, following that of similarly-structured methane. We use fixed values of each species' heat capacity for simplicity; we verify that temperature-dependent values from NIST \citep{NIST} deviate by less than a factor of two from these constant values over the adiabatic region, which only marginally alters the atmospheric structure we obtain. Generally, we find that the atmospheric composition in the adiabatic regime is hydrogen-dominated, so the heat capacity is very near that of hydrogen alone. We assume the hydrogen remains fully diatomic throughout the atmosphere. Hydrogen dissociation is expected at sufficiently high pressures and temperatures. Molecular dynamics simulations indicate that hydrogen likely remains diatomic throughout most of our atmospheric structure, though it may start to dissociate at the highest temperatures and pressures we consider \citep[e.g.][]{TamblynBonev2010, FrenchBecker2012, SoubiranMilitzer2017}. If hydrogen begins to dissociate, the effective adiabatic index of the atmosphere would be lower due to the energy required for breaking the H--H bonds \citep{LC15}, similar to the effect of latent heat of vaporization, making the adiabatic temperature gradient shallower.

To construct the atmospheric structure, we begin from the outer radiative-convective boundary, where $R=R_\mathrm{rcb}$ and $P(R_\mathrm{rcb})$ are chosen, and $T(R_\mathrm{rcb})=T_\mathrm{eq}$. We increment in small pressure steps: $P_\mathrm{new} = P + \Delta P$. The new radius is thus $R_\mathrm{new} = R + \Delta P/(\partial P/\partial R)$, and the new temperature, $T_\mathrm{new} = T + \Delta P (\partial T/\partial P)$, employing Eqs.~\ref{eq:dPdr} and \ref{eq:wetadiabat} respectively.

As described in \citet{MS22}, the change in composition of the atmosphere with temperature due to condensation causes a mean molecular weight gradient. In a hydrogen-dominated atmosphere, these condensation effects cause the mean molecular weight to increase with temperature, stabilizing the gas against convection. This is in contrast to an Earth-like atmosphere, in which the major condensable, water, is lighter than the background air. In previous work, in which one condensable species was considered, the tipping point beyond which convection is inhibited could be quantified by a critical mass mixing ratio $q_\mathrm{crit}$ \citep{G95, L17, MS22, Markham22}: 
\begin{equation}\label{eq:qcrit}
    q_\mathrm{crit} = \cfrac{1}{\bigg(1-\cfrac{\mu_\mathrm{H}}{\mu_\mathrm{cond}}\bigg) \cfrac{\partial \ln P_\mathrm{cond}}{\partial \ln T}}.
\end{equation}
In an atmosphere with multiple species changing abundance, this relationship is in principle more complex. However, the inhibition of convection can be quantified by calculating a "convection criterion" for each non-hydrogen species $\chi_i$:
\begin{equation}\label{eq:conv_crit}
     \chi_i = q_i \bigg(1-\frac{\mu_\mathrm{H}}{\mu_\mathrm{i}}\bigg) \frac{\partial \ln P_\mathrm{i}}{\partial \ln T},
\end{equation}
where $q_i$ is the mass-mixing ratio of species $i$. Convection is inhibited if $\sum_i \chi_i \geq 1$. In the case of a single species, this method yields the same results as Eq.~\ref{eq:qcrit}. We derive the fact that the overall stability criterion is the sum of the criteria for each species in Appendix~\ref{sec:criterion}.

\subsubsection{Non-convective region}\label{sec:nonconv}
In regions where the criterion is fulfilled, convection is inhibited, no matter how super-adiabatic the temperature profile becomes. In these regions, heat must be transported by either radiation or conduction. At the typical temperatures and pressures in the deep non-convective regions of sub-Neptunes we consider in this work, conduction may be competitive with radiation in transporting heat \citep[e.g.][]{Vazan2020, MS22}. 

We quantify the competition between conduction and radiation by comparing the conductivity, $\lambda_\mathrm{cond} = L/(4 \pi r^2)/(\partial T/\partial r)_\mathrm{cond}$, to the equivalent radiative heat transport term, $\lambda_\mathrm{rad} \equiv L/(4 \pi r^2)/(\partial T/\partial r)_\mathrm{rad} = 16 \sigma T^3/(3 \kappa \rho)$, where where $\sigma$ is the Stefan-Boltzmann constant and $\kappa$ is the local Rosseland mean opacity. Conduction dominates radiation when $\Lambda \equiv \lambda_\mathrm{cond}/\lambda_\mathrm{rad} > 1$, or, written in scaling form:
\begin{equation}\label{eq:condrad}
\begin{split}
    \Lambda &\approx \bigg(\frac{\lambda_\mathrm{cond}}{7 \times 10^4~\mathrm{erg~s^{-1} cm^{-1} K^{-1}}}\bigg) \bigg(\frac{\kappa}{10^3~\mathrm{cm^2 g^{-1}}}\bigg) \\ 
    & \bigg(\frac{\rho}{1~\mathrm{g~cm^{-3}}}\bigg) \bigg(\frac{T}{6000~\mathrm{K}}\bigg)^{-3} \\
    &> 1 \mathrm{.}
\end{split}
\end{equation}
From equation~(\ref{eq:condrad}), it is apparent that larger conductivities, opacities, and gas densities favor conduction over radiation.

Due to the widely varying atmospheric compositions, the atmospheric Rosseland mean opacity $\kappa$ is difficult to determine without a detailed radiative model, which is beyond the scope of this work.
Therefore, following \citet{MS22} and \citet{Markham22}, we use the \citet{F14} relation for solar metallicity gas throughout the atmosphere. We extend these opacities to pressures beyond their asserted validity, so we acknowledge that the opacity of the interior is a source of uncertainty in our model. We discuss the relevance of different opacity choices in Section~\ref{sec:results}.

As with the opacities, the conductivities in these regions are uncertain and depend on the material properties of the atmosphere, which are not entirely clear for the exotic mixtures we encounter. Therefore, we employ a simple approach based on the behavior of pure H/He, as described in \citet{MS22}. To calculate the thermal conductivity, we use the electrical conductivity scaling with temperature and pressure from \citet{McWilliams2016}, which is based on experimental results for pure hydrogen. We then convert these to thermal conductivities using the Wiedemann-Franz law. At relatively low temperatures and pressures, where the electrical conductivity is low, we use a minimum value of $\lambda_\mathrm{cond} = 2 \times 10^5$ erg s$^{-1}$ cm$^{-1}$ K$^{-1}$, appropriate for the nucleic contribution over a broad range of relevant temperatures and pressures \citep{FrenchBecker2012} and consistent with approximations used in previous work modeling Earth-like silicate in planetary interiors \citep[e.g.][]{Stevenson1983, Vazan2020}.

We can incorporate both conduction and radiation by adding the thermal conductivity to the equivalent radiative term to make an ``effective conductivity'', $\lambda_\mathrm{eff} \equiv
\lambda_\mathrm{cond} + \lambda_\mathrm{rad}$. An alternative, but equivalent, framing is to employ an effective opacity, $\kappa_\mathrm{eff}$ \citep{Vazan2020}:

\begin{equation}
    \kappa_\mathrm{eff} = \cfrac{1}{\cfrac{1}{\kappa} + \cfrac{1}{\kappa_\mathrm{cond}}}
\end{equation}
where $\kappa_\mathrm{cond}$, the ``conductive opacity'', is given by
\begin{equation}
    \kappa_\mathrm{cond} = \frac{16 \sigma T^3}{3 \rho \lambda_\mathrm{cond}}.
\end{equation}

The temperature gradient of the non-convective region is then determined by the energy flux, $L$, as well as the local pressure, temperature, and the effective opacity $\kappa_\mathrm{eff}$:
\begin{equation}\label{eq:radtemp}
    \frac{\partial \ln T}{\partial \ln P} = \frac{3 \kappa_\mathrm{eff} P L}{64\pi G M_\mathrm{c} \sigma T^4}.
\end{equation}
In steady state, the energy flux that must be transported across the radiative boundary is equal to the radiative flux of the planet into space, i.e., the luminosity at the radiative-convective boundary:
\begin{equation}\label{eq:luminosity}
    L = \frac{\gamma-1}{\gamma}\frac{64\pi G M_\mathrm{c} \sigma T_\mathrm{eq}^4}{3 \kappa_\mathrm{rcb} P_\mathrm{rcb}},
\end{equation}
where `rcb' subscripts represent values calculated at the conditions of the radiative-convective boundary. Here, the atmosphere is dominated by hydrogen, so we use the adiabatic index for pure H$_2$ gas and opacity relations of \citet{F14} for solar composition gas.

\section{Results}\label{sec:results}
In this section, we present the atmospheric profile and chemical abundances we obtain for our fiducial planet, a 4 $M_\oplus$ planet with an equilibrium temperature of 1000~K. The atmosphere has a total hydrogen mass of 2.5\% the core's total mass. These values are all typical of sub-Neptunes \citep[e.g.][]{LF14}. This planet has a base temperature of 5000~K, which is a reasonable value early in the planet's evolution \citep[e.g.][]{GSS16, MS22}. We begin by examining the region interior to the outer radiative-convective boundary in Section \ref{sec:inner}, where the chemistry has the largest effect on the atmospheric structure. We then describe the chemical equilibrium of the outer radiative region in Section \ref{sec:outer}, which has implications for the observability of these interior-atmosphere interactions.

\subsection{Inner atmosphere}\label{sec:inner}
\begin{figure}
	\includegraphics[width=\columnwidth]{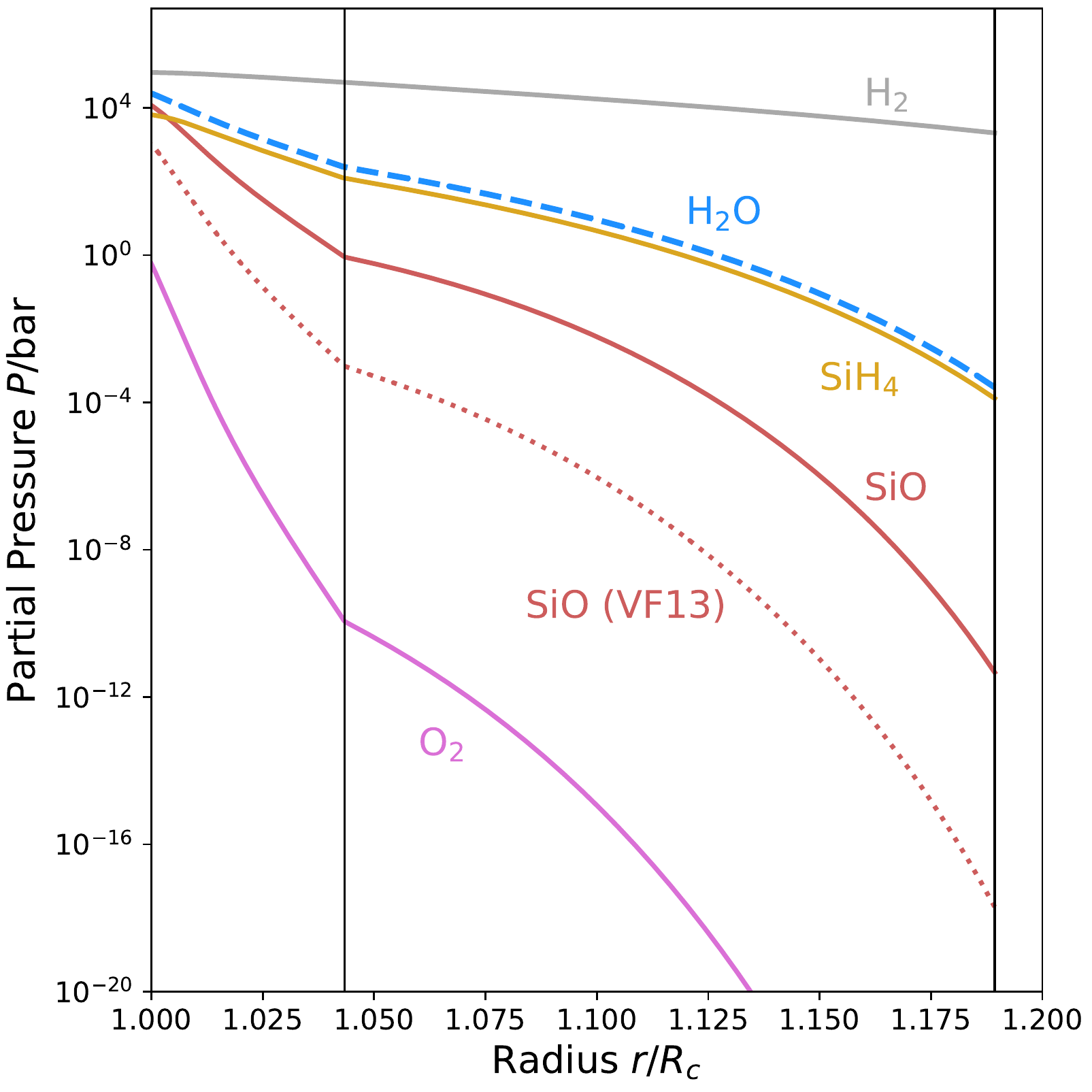}
    \caption{Partial pressures of the chemical species present in the atmosphere as a function of radius, in core radii $R_\mathrm{c}$, for a 4~$M_\oplus$ planet with an equilibrium temperature of 1000~K, a base temperature of 5000~K, and an atmospheric hydrogen mass fraction of 2.5\%. The species we consider are H$_2$ (gray), H$_2$O (blue dashed), SiH$_4$ (yellow), SiO (red solid), and O$_2$ (pink). For comparison, the SiO abundance derived from the congruent evaporation equation in \citet{VF13} is shown as a red dotted line. The outer vertical black line at 1.19 $R_\mathrm{c}$ represents the outer radiative convective boundary, while the inner black line at 1.04 $R_\mathrm{c}$ represents the inner transition within which convection is inhibited.}
    \label{fig:chem}
\end{figure}

\begin{figure}
	\includegraphics[width=\columnwidth]{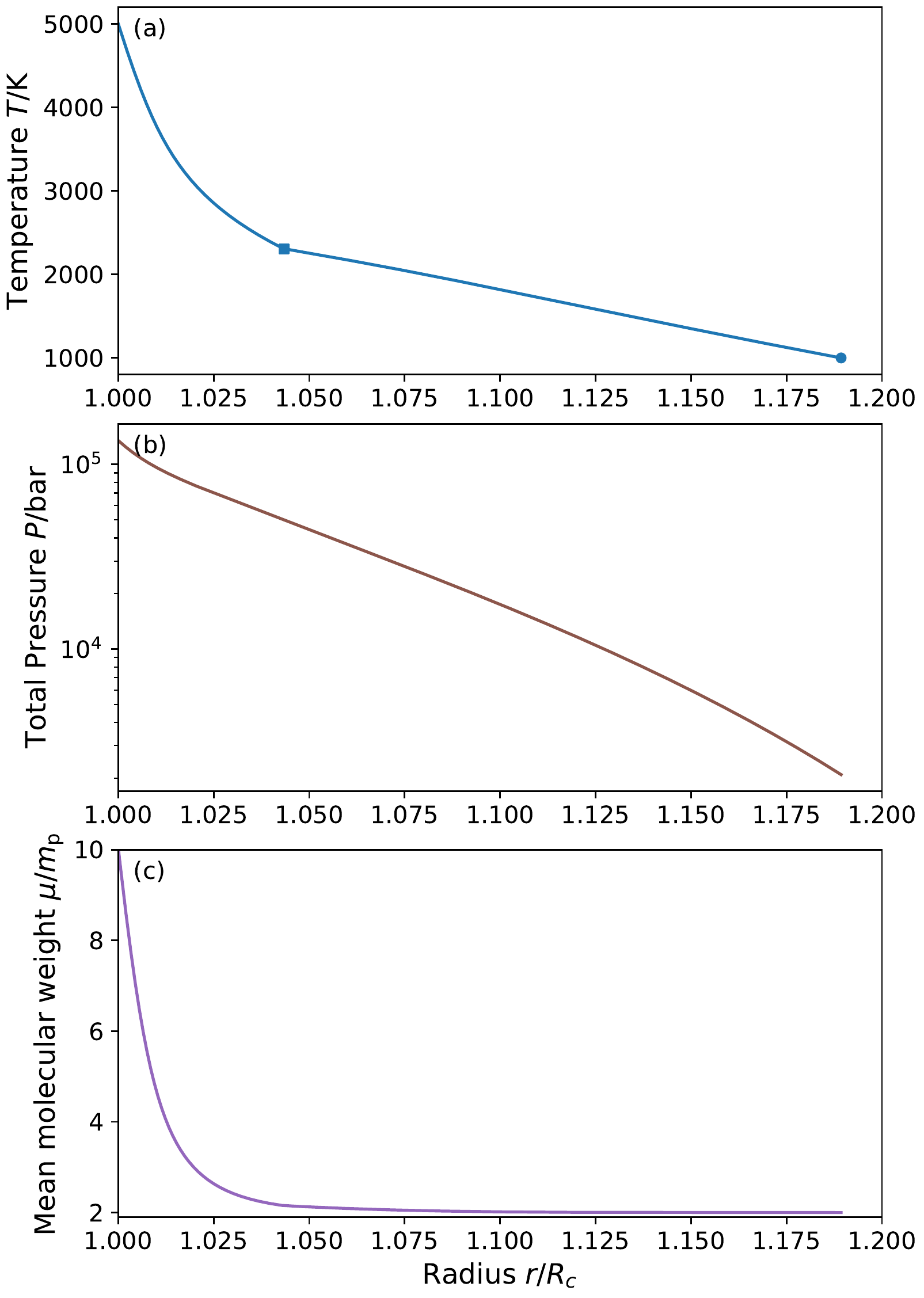}
    \caption{Example of sub-Neptune envelope structure. Panel (a) shows the temperature $T$ in kelvin, (b) the total pressure $P$ in bar, and (c) the mean molecular weight $\mu$ in proton masses $m_\mathrm{p}$, as functions of radius, in core radii $R_\mathrm{c}$. The model is the same one as shown in Fig.~\ref{fig:chem}: a 4~$M_\oplus$ planet with an equilibrium temperature of 1000~K, a base temperature of 5000~K, and an atmospheric hydrogen mass fraction of 2.5\%. In the top panel, the dot represents the outer radiative-convective boundary, while the square represents the inner transition, inside of which convection is inhibited.}
    \label{fig:prof}
\end{figure}

\begin{figure}
	\includegraphics[width=\columnwidth]{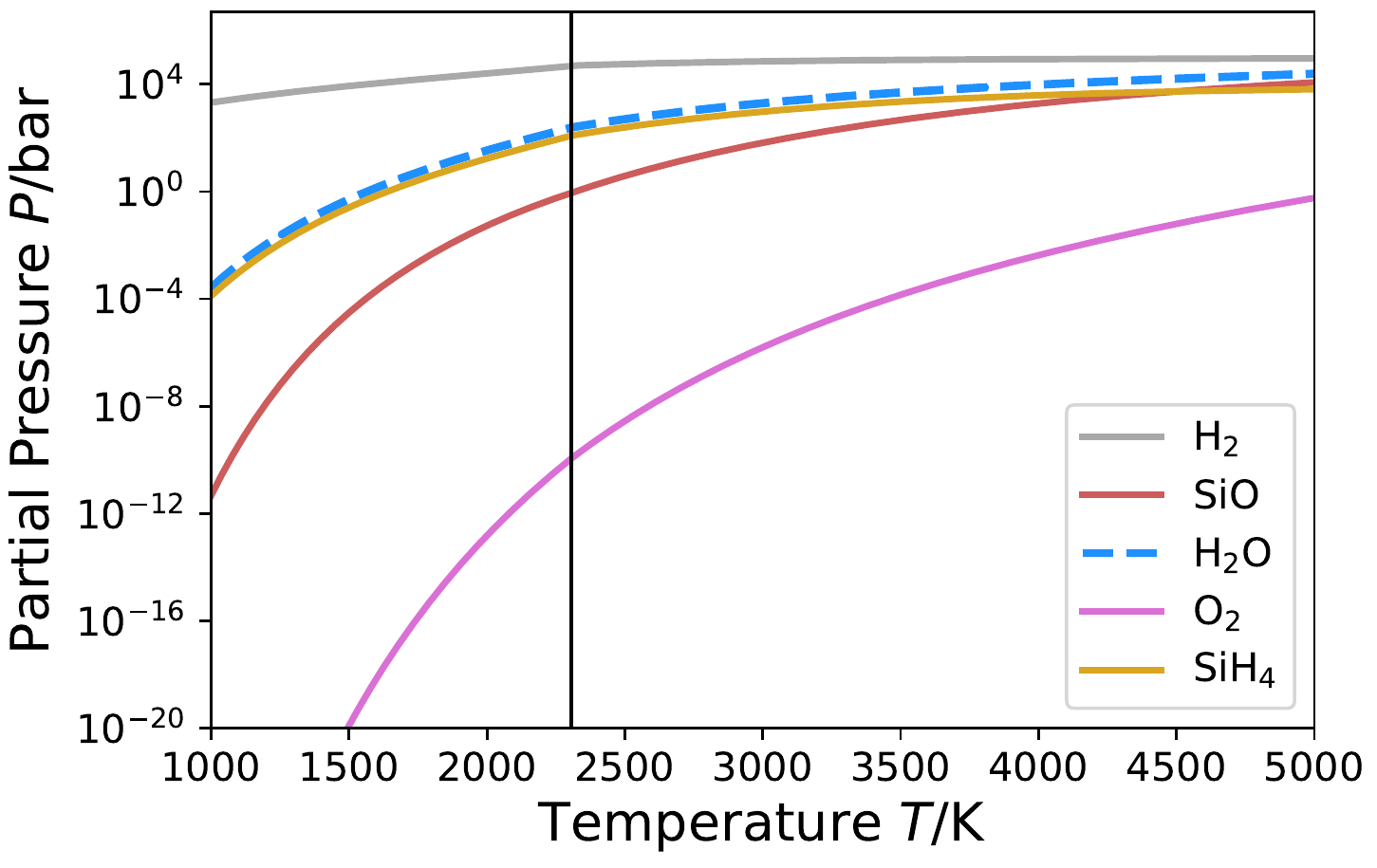}
    \caption{Partial pressures of the chemical species present in the atmosphere as a function of temperature. The line meanings remain the same as those in Fig.~\ref{fig:chem}, as do the physical parameters: a 4~$M_\oplus$ planet with an equilibrium temperature of 1000~K, a base temperature of 5000~K, and an atmospheric hydrogen mass fraction of 2.5\%. The black line at 2300~K represents the point at which convection becomes inhibited at hotter temperatures.}
    \label{fig:PT}
\end{figure}

In this section we focus on the region interior to the outer radiative-convective boundary, which is at $\sim 1.19 R_\mathrm{c}$ in our fiducial model. This radius and corresponding pressure $P(R_\mathrm{rcb})$ are found by iterating the atmospheric profile until the desired planet characteristics, i.e. the hydrogen mass and base temperature $T(R_\mathrm{c})$, are achieved, following the method of \citet{MS22}. In Fig.~\ref{fig:chem} we show the partial pressures of the species we consider, namely H$_2$ (gray), H$_2$O (blue dashed), SiH$_4$ (yellow), SiO (red solid), and O$_2$ (pink), as functions of radius. In Fig.~\ref{fig:prof}, we show aspects of the overall atmospheric profile, namely the temperature, pressure, and mean molecular weight as functions of radius, and in Fig.~\ref{fig:PT} we present an alternative view of the same model as in Fig.~\ref{fig:chem}, but in pressure-temperature space.

\begin{figure}
	\includegraphics[width=\columnwidth]{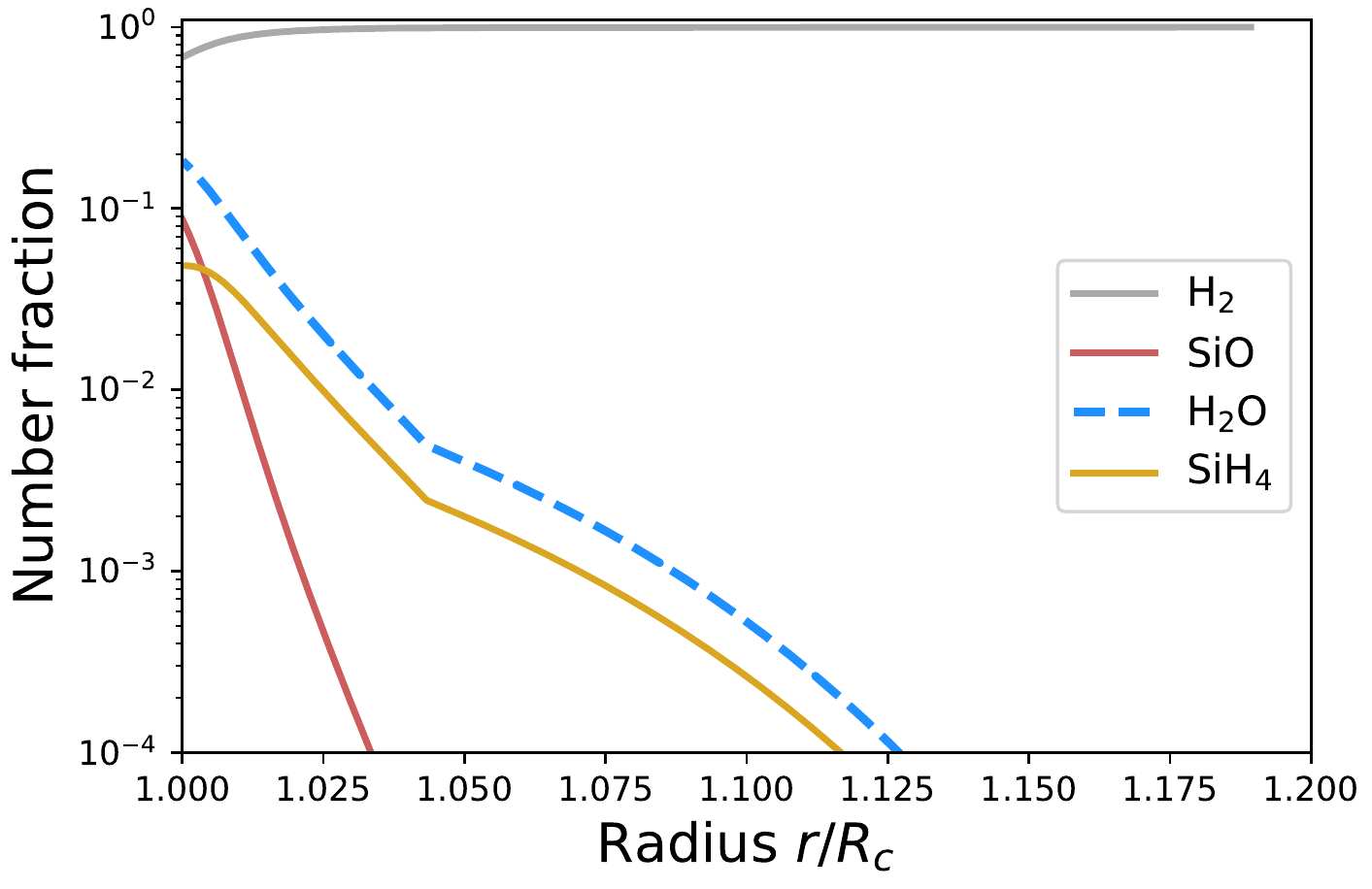}
    \caption{The number fraction of the species we consider as a function of radius, in core radii $R_\mathrm{c}$. The line meanings remain the same as those in Fig.~\ref{fig:chem}, as do the physical parameters: a 4~$M_\oplus$ planet with an equilibrium temperature of 1000~K, a base temperature of 5000~K, and an atmospheric hydrogen mass fraction of 2.5\%. The number fraction of oxygen remains less than $10^{-4}$ and so is not shown. The atmosphere remains hydrogen dominated by number.}
    \label{fig:num_frac}
\end{figure}

\begin{figure}
	\includegraphics[width=\columnwidth]{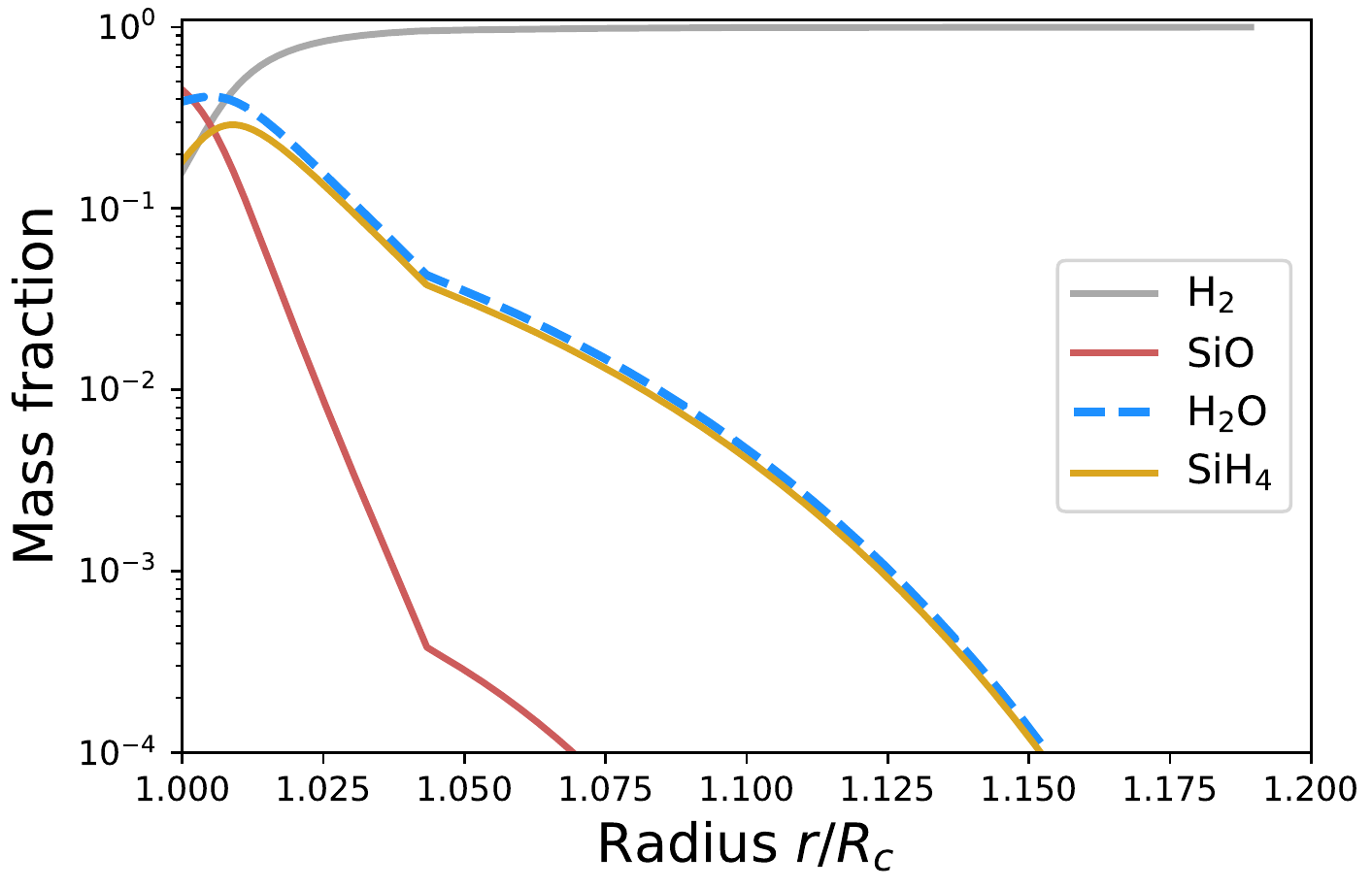}
    \caption{The mass fraction of the different species as a function of radius, in core radii $R_\mathrm{c}$. The line meanings remain the same as those in Fig.~\ref{fig:chem}, as do the physical parameters: a 4~$M_\oplus$ planet with an equilibrium temperature of 1000~K, a base temperature of 5000~K, and an atmospheric hydrogen mass fraction of 2.5\%. The mass fraction of oxygen remains less than $10^{-4}$ and so is not shown. Near the base of the atmosphere, when $T \sim 5000$~K, the atmosphere becomes dominated in mass by water and SiO.}
    \label{fig:mass_frac}
\end{figure}

The atmosphere is hydrogen-dominated by number at all radii (i.e., $P_\mathrm{H_2} > P_i$ for all other species $i$). The dominance of hydrogen by number is demonstrated in Fig.~\ref{fig:num_frac}, which shows the number fraction of the species we consider throughout the atmosphere. The abundances of all the secondary species we consider, namely H$_2$O, SiH$_4$, SiO, and O$_2$, increase in abundance with depth as the temperature increases. Water and silane have partial pressures a factor of $\sim 10^{-7}$ lower than that of H$_2$ at the outer radiative-convective boundary, while SiO is lower still, with a number fraction of $\sim 10^{-14}$, and O$_2$ many orders of magnitude less than this. However, at the magma-atmosphere interface, $r = R_\mathrm{c}$, secondary species abundances are much higher, with SiO, H$_2$O, and SiH$_4$ together making up nearly 40\% of the atmosphere by number and overtaking H$_2$ by mass, as shown in Fig.~\ref{fig:mass_frac}. Accordingly, the mean molecular weight, displayed in panel (c) of Fig.~\ref{fig:prof}, remains near 2 $m_\mathrm{p}$, that of H$_2$, throughout most of the atmosphere but rises sharply to larger than 8 $m_\mathrm{p}$ near the inner edge.

\begin{figure}
	\includegraphics[width=\columnwidth]{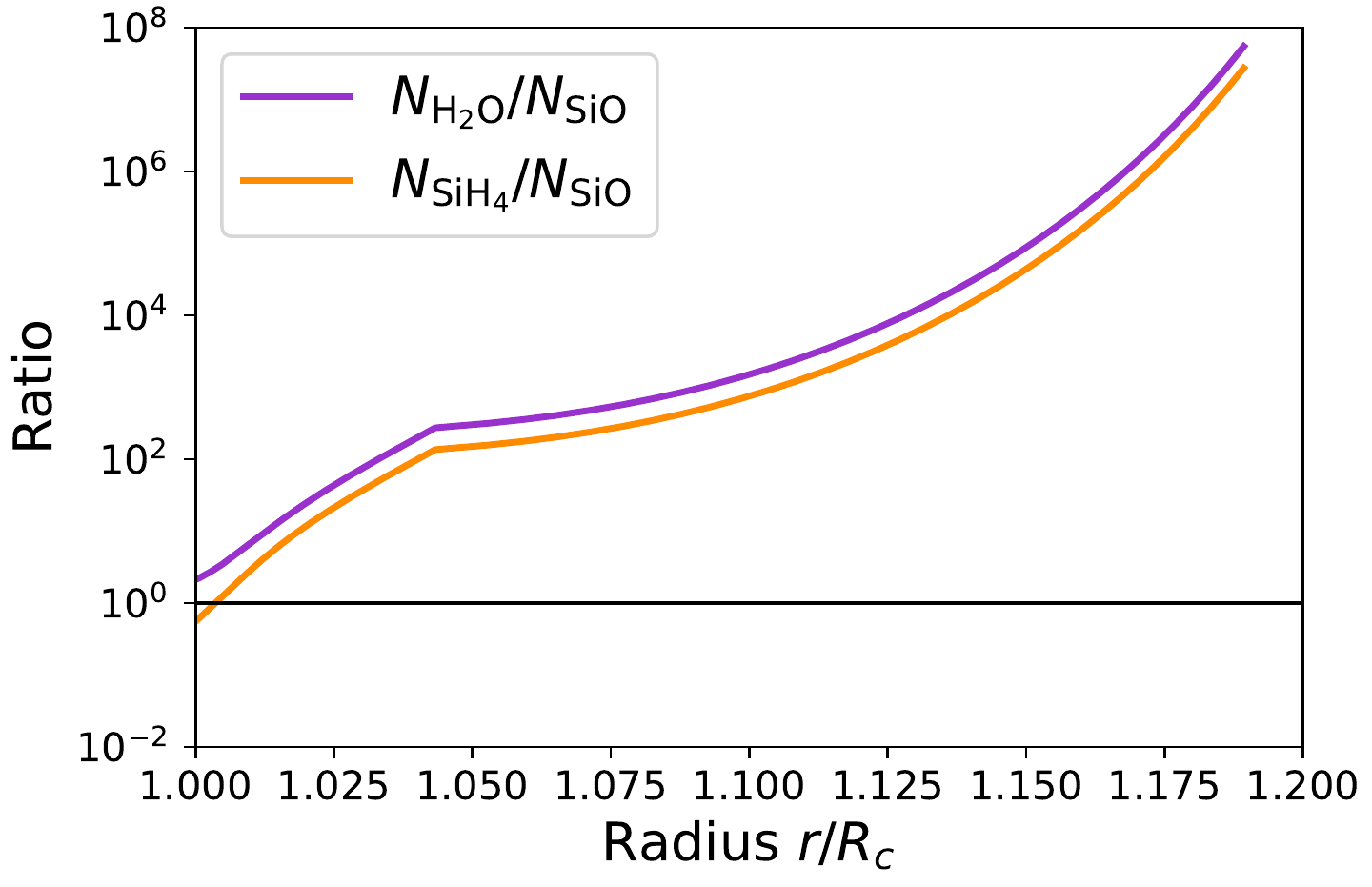}
    \caption{Number ratios of H$_2$O to SiO (in purple) and SiH$_4$ to SiO (in orange) as a function of radius, in core radii $R_\mathrm{c}$. The model is the same one as shown in Fig.~\ref{fig:chem}: a 4~$M_\oplus$ planet with an equilibrium temperature of 1000~K, a base temperature of 5000~K, and an atmospheric hydrogen mass fraction of 2.5\%. The more reduced species, H$_2$O and SiH$_4$, dominate at nearly all points in the atmosphere, except near the hot base.}
    \label{fig:ratios}
\end{figure}

We show the number ratios of H$_2$O to SiO (in purple) and SiH$_4$ to SiO (in orange) in Fig.~\ref{fig:ratios}. Throughout most of the atmosphere, SiH$_4$ dominates over SiO, except at the very base for our chosen parameters, and H$_2$O dominates SiO in the entire atmosphere. This dominance of reduced species compared to oxidized ones is due to the presence of hydrogen gas as the background species. In contrast to inert background gases such as nitrogen, hydrogen is highly reactive and will act to reduce the outgassed magma ocean species, per Reactions~\ref{eq:O2rxn} and \ref{eq:SiH4rxn}. We find these equations are typically driven toward the products at the temperatures and hydrogen gas fractions we consider, favoring the production of water and silane as the primary oxygen and silicon carriers in the atmosphere, respectively. This is in contrast to previous work on the atmosphere-interior interaction, such as \citet{MS22} and \citet{Markham22}, which considered only SiO as the main carrier of the silicon and oxygen taken up from the underlying magma ocean.

Due to the numbers of silicon and oxygen atoms remaining in a fixed ratio, as described by Eq.~\ref{eq:Nrelation}, the partial pressures of water and silane are similar, and remain in lockstep as the temperatures increase. At sufficiently high temperatures deep in the atmosphere, SiH$_4$ production is no longer favored over SiO, and the abundance of SiO approaches and, for the conditions we consider, just surpasses that of SiH$_4$, as shown most clearly in Fig.~\ref{fig:num_frac}. This behavior appears qualitatively similar to the CH$_4$--CO transition, extensively studied in the context of exoplanet observations \citep[e.g.][]{BurrowsSharp1999, VisscherMoses2010, FortneyVisscher2020}, though that transition occurs at much lower temperatures (see also Section \ref{sec:outer} below).

Another notable result shown by Fig.~\ref{fig:chem} is that the overall abundance of silicon-bearing species is much higher than was previously found in \citet{MS22}. In that work, the authors used a vapor pressure formula appropriate for congruent evaporation of silicate magma into a vacuum taken from \citet{VF13}. This value for rock vapor is denoted in Fig.~\ref{fig:chem} with a red dotted line. However, \citet{MS22} ignored the chemical effects of the non-inert background gas, H$_2$. When we account for the hydrogen chemistry, we find that the production of silane and water via Reaction~\ref{eq:SiH4rxn} draws out more SiO from the interior in order to continue to satisfy the relationship with $K_\mathrm{eq, R3}$, which is solely a function of temperature, in Equation~\ref{eq:SiH4K}. In order to maintain the equality of Equation~\ref{eq:SiOK}, the oxygen partial pressure decreases to the values in Fig.~\ref{fig:chem}. These values are much lower than the congruent ratio of 0.5 mole of O$_2$ for every mole of SiO.

Meanwhile, water is the most abundant product of magma-hydrogen interaction in our model, despite no water being present in the system initially. Water as a fundamental byproduct of silicate-hydrogen chemistry has previously been found in the context of sub-Neptunes \citep{SY22, Zilinskas2023} and early Earth \citep{YoungShahar2023}. This result shows that the presence of water in a sub-Neptune atmosphere does not necessarily indicate it formed with substantial ices, as has been suggested for some sub-Neptunes \citep[e.g.][]{Zeng2019, Venturini20, Madhusudhan20, Emsenhuber21}. We investigate factors that could alter the abundance of water in a sub-Neptune in Section \ref{sec:otherspecies}. 

The increased abundances of condensable species alters the overall atmospheric structure compared to the previous SiO-only model of \citet{MS22}. These increased abundances increase the value of the convective criterion, and therefore lower the temperature at which the atmosphere transitions from convective to radiative to $\sim 2300$~K, as marked in Fig.~\ref{fig:prof} by the square, a temperature significantly lower than the typical transition temperature of $\sim 4000$~K found in \citet{MS22} for similar planet parameters.

\begin{figure}
	\includegraphics[width=\columnwidth]{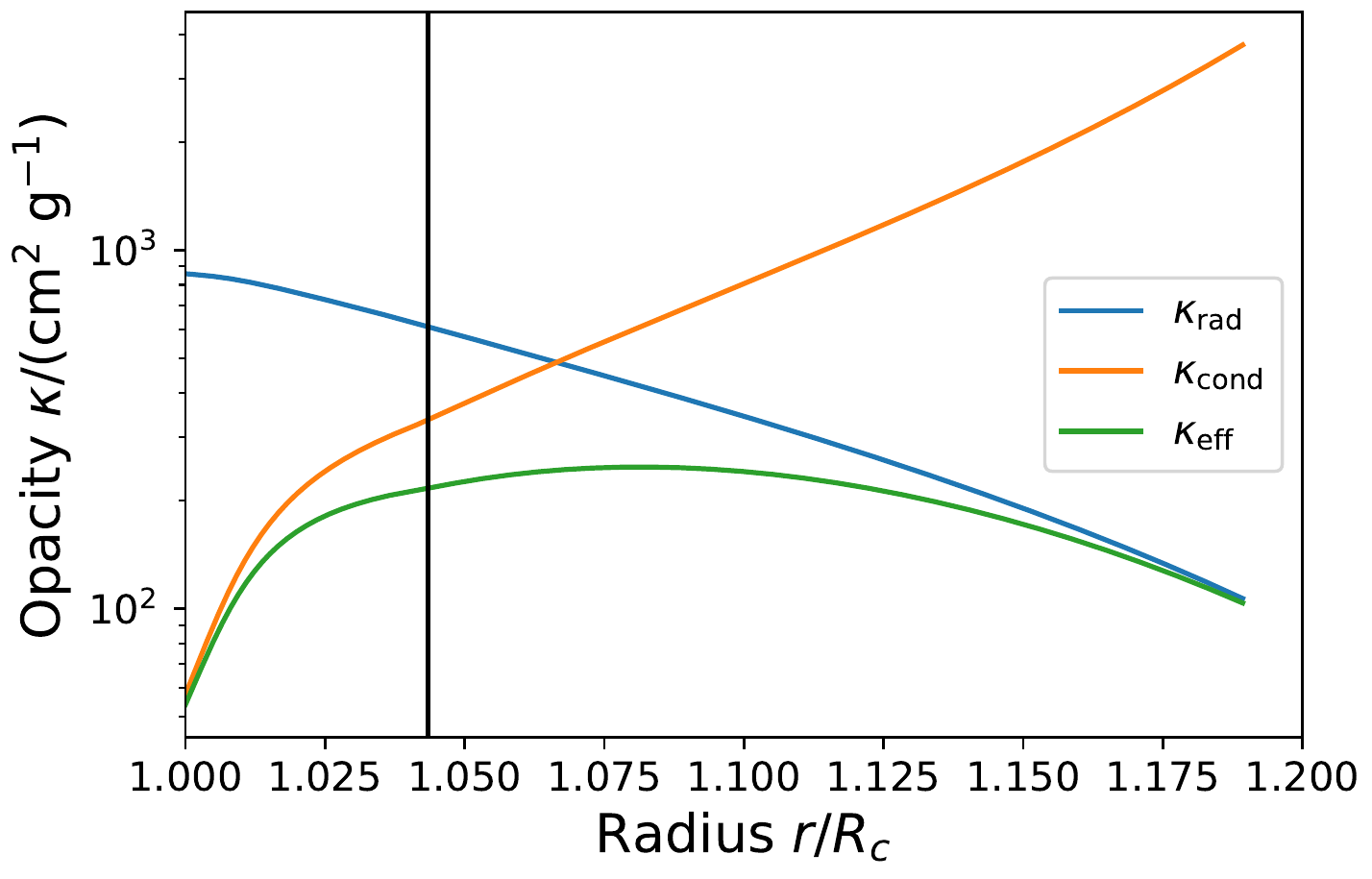}
    \caption{Comparison of the radiative opacity $\kappa_\mathrm{rad}$ (in blue), the conductive opacity $\kappa_\mathrm{cond}$ (in orange), and the effective opacity $\kappa_\mathrm{eff}$ (in green), as a function of radius, in core radii $R_\mathrm{c}$. The model is the same one as shown in Fig.~\ref{fig:chem}: a 4~$M_\oplus$ planet with an equilibrium temperature of 1000~K, a base temperature of 5000~K, and an atmospheric hydrogen mass fraction of 2.5\%. The black line represents the inner non-convective boundary. Radiation dominates at the outer radiative-convective boundary, as expected, while conduction becomes more important than radiation in the interior, though by a factor of 10 or less.}
    \label{fig:condrad}
\end{figure}

A transition at lower temperatures means the atmosphere is non-convective for a larger range of temperatures in the deep interior of sub-Neptune planets, as can be seen in Fig.~\ref{fig:prof}. The gas density at the transition point is also lower. Since the opacities and conductivities we expect are lower at lower densities, the temperature-pressure profiles are not as steep in the non-convective region as was found in \citet{MS22}. We compare the opacities we use in Fig.~\ref{fig:condrad}. We find that radiation and conduction are comparably efficient in transporting energy deep in sub-Neptune atmospheres, as found in \citet{MS22}, with conduction dominating as the temperatures and pressures increase. As discussed in Section~\ref{sec:nonconv}, the opacity we use is extrapolated beyond the pressure regime fit in \citet{F14}. However, it is apparent in Fig.~\ref{fig:condrad} that the effective opacity is close to the conductive opacity in the interior, with the radiative opacity being larger. If the opacity were larger than the solar value we use ($\sim 10^3$ cm$^2$ g$^{-1}$), as might be expected in a hot, high-metallicity region, heat transport would be even more conduction-dominated than we find here, with little effect on the structure we obtain. We therefore conclude that our results are insensitive to the exact value of the opacity in the interior so long as it is higher than the conductive opacity. This is similar to the conclusion reached regarding opacities in \citet{MS22}. Fig.~\ref{fig:condrad} also confirms that radiation dominates conduction at the outer radiative-convective boundary.

\subsection{Outer atmosphere}\label{sec:outer}
\begin{figure}
	\includegraphics[width=\columnwidth]{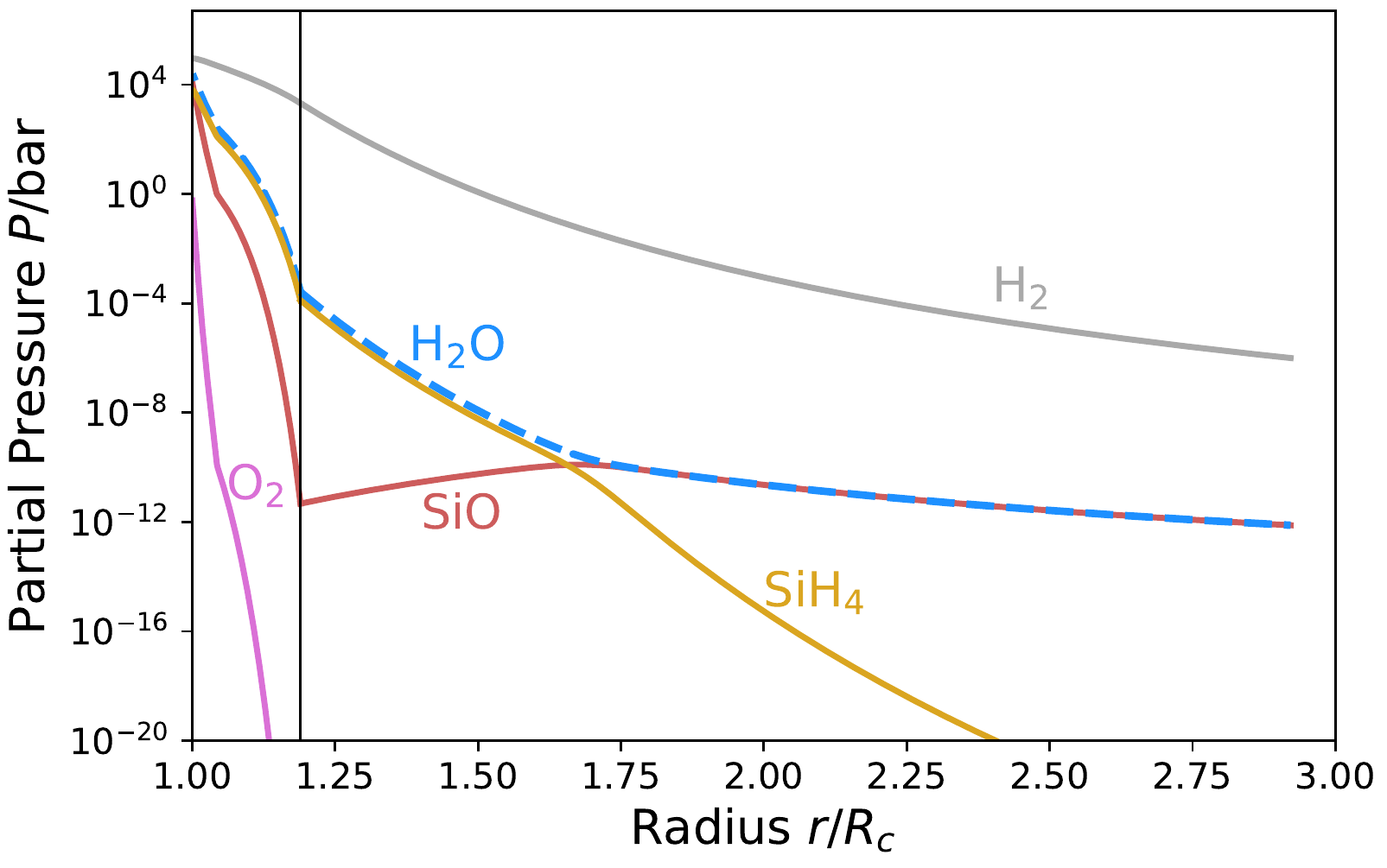}
    \caption{Partial pressures of the chemical species present in the atmosphere as a function of radius, in core radii $R_\mathrm{c}$, through the entire atmosphere out to $P = 10^{-6}$~bar ($r \sim 3 R_\mathrm{c}$), including the outer radiative region. The profile is for a 4~$M_\oplus$ planet with an equilibrium temperature of 1000~K, a base temperature of 5000~K, and an atmospheric hydrogen mass fraction of 2.5\%. The species we consider are H$_2$ (gray), H$_2$O (blue dashed), SiH$_4$ (yellow), SiO (red solid), and O$_2$ (pink). The outer vertical black line at 1.19 $R_\mathrm{c}$ represents the outer radiative convective boundary.}
    \label{fig:total_chem}
\end{figure}
We now examine the implications of this interior chemistry and structure on the outer atmosphere, the region accessible to spectroscopic observations. We model the outer atmosphere as isothermal at $T=T_\mathrm{eq}$, which for our model planet is set to 1000~K. In this region, the total pressure falls off exponentially. We assume chemical equilibrium is maintained throughout the atmosphere and that the vapor is saturated in SiO$_2$, with the SiO$_2$ activity always equal to one. In particular, by assuming chemical equilibrium we implicitly assume that the region we probe is below the homopause, above which each species follows its own scale height, and that mixing is always faster than the chemical kinetic timescale, i.e., the reactions we consider are never quenched. In Fig.~\ref{fig:total_chem} we show the partial pressures of the chemical species we consider as a function of planet radius over the whole atmosphere. The interior region, $r < R_\mathrm{rcb} \approx 1.19 R_\mathrm{c}$, contains the same abundances presented in Fig.~\ref{fig:chem}.

\begin{figure}
	\includegraphics[width=\columnwidth]{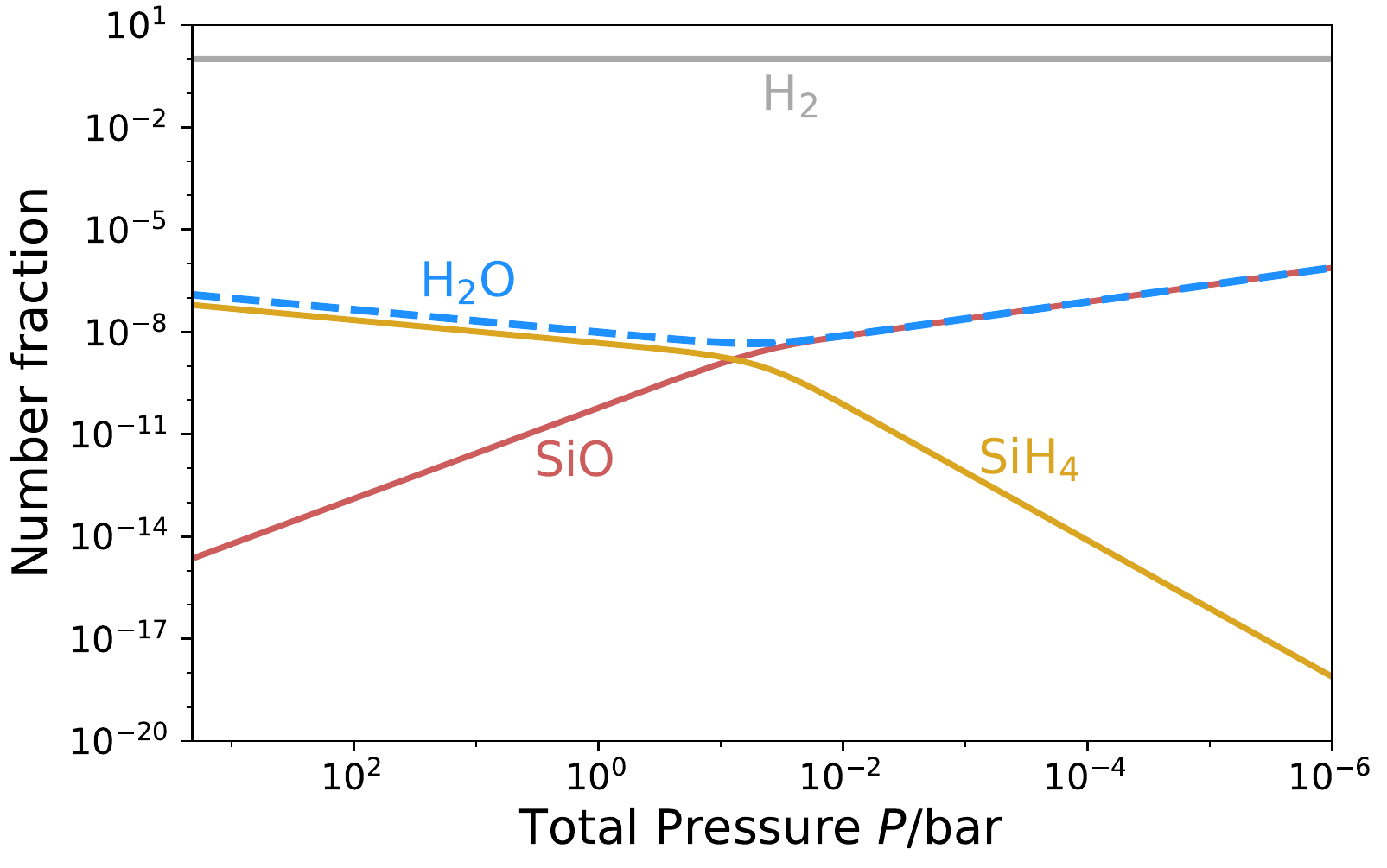}
    \caption{Number fraction of each chemical species as a function of total pressure in the outer isothermal region. The temperature is fixed at 1000~K. The number fraction of O$_2$ remains less than $10^{-20}$ and so is not shown. SiO becomes the dominant silicon-bearing species over SiH$_4$ at $\sim 10^{-1}$~bar.}
    \label{fig:NvsP}
\end{figure}

Fig.~\ref{fig:total_chem} shows sharp kinks in the abundances of the species we consider as the temperature ceases declining. Intriguingly, not only do the abundances change with altitude in the isothermal region, but the relative proportions of the different species do as well. The changes in relative abundances is shown more clearly in Fig.~\ref{fig:NvsP}, which displays the number fraction of each species as a function of the total pressure in the isothermal region only. This total pressure is very nearly the pressure of H$_2$, as evidenced by its number fraction being nearly 1 in this region. The relative abundance of SiO increases throughout the isothermal region, while the relative abundance of SiH$_4$ decreases. The two become equal near $10^{-1}$~bar: at lower pressures (i.e. higher altitudes) than this, SiO is the dominant silicon-bearing species, rather than SiH$_4$.

The increasing dominance of SiO over SiH$_4$ can be understood by examining Eq.~\ref{eq:SiH4K}. At constant temperature, $K_\mathrm{eq, R3}$ is constant. As $P_\mathrm{H_2}$ decreases with altitude, the partial pressure of SiO must increase to maintain the equilibrium. In other words, the reaction no longer so strongly favors the production of SiH$_4$. Due to the fixed Si:O atomic number ratio, Eq.~\ref{eq:Nrelation}, the abundance of water first decreases with decreasing pressure, tied to the abundance of SiH$_4$, then increases in lockstep with SiO once the latter becomes dominant.

This transition from SiH$_4$ to SiO as the dominant silicon-bearing species occurs at an observationally relevant pressure for the planet parameters studied here. Therefore, the relative SiO/SiH$_4$ abundance could serve as a probe of the overall chemistry of the interior and atmosphere, similarly to the well-studied CO--CH$_4$ transition, which occurs due to a similar reaction \citep[e.g.][]{BurrowsSharp1999, VisscherMoses2010, FortneyVisscher2020} . Future work will elucidate how this transition point varies across the sub-Neptune parameter space and in time, as well as with more complex interior compositions. It will also further constrain the absolute abundances we expect. We note that our simple model assumes the initial atmosphere comprises pure hydrogen, such that all heavier species in our results are due to outgassing from the magma interior. This results in elemental abundances in the upper atmosphere which are sub-solar, and which do not match solar ratios. Specifically, the solar abundance of silicon is $3.3 \times 10^{-5} N_\mathrm{H}$, and that of oxygen is $5.7 \times 10^{-4} N_\mathrm{H}$, where $N_\mathrm{H}$ is the hydrogen abundance \citep{Lodders2021}, implying an oxygen-to-silicon ratio of $N_\mathrm{O}/N_\mathrm{Si} = 17.3$. Our abundances and oxygen-to-silicon ratio are lower than these values throughout the isothermal region. We further discuss the effects of different compositions of core and atmosphere on our results in Section \ref{sec:otherspecies}.

\section{Discussion and Future Directions}
In Sec.~\ref{sec:results}, we demonstrated that silane and water are expected to be byproducts of silicate-hydrogen interactions in sub-Neptunes with underlying magma oceans. In this section, we discuss the observability of these species with current and future facilities. We then discuss how other species beyond the simplified chemical network we consider here could alter the atmospheric profiles we obtain, how these new profiles could alter the evolution in time of these atmospheres, and how our assumptions about opacities could impact these results.

\subsection{Observability}
The two most abundant species in our model besides hydrogen are water and silane. Water vapor has numerous strong absorption bands in the infrared, which JWST is well-suited to exploit. Due to its potential as a tracer of planet formation and its importance for life, water vapor has been extensively searched for in the atmospheres of sub-Neptunes in previous campaigns with Hubble and JWST. Water features have been detected in Hubble observations of sub-Neptunes \citep{Benneke19GJ3470, Benneke19K218} and in the atmospheres of giant exoplanets using JWST \citep[e.g.][]{WASP39NIRSpec, WASP39PRISM}, with observations of smaller planets underway. Our results show that detections of water vapor in sub-Neptune atmospheres is not necessarily diagnostic of formation beyond the snow-line. Instead, it could be produced endogenously via magma-hydrogen interactions. 

Silane also has features in the mid-infrared, with the largest cross-section at $\sim 4.5\ \mu$m \citep{Owens17ExoMolSiH4} according to the ExoMol database \citep{ExoMol}. This feature is squarely within the wavelength capabilities of JWST's NIRSpec and PRISM instruments, as evidenced by the detections of  features at similar wavelengths, such as CO$_2$ in the hot Jupiter WASP-39 b \citep{JWSTCO2}. Silane has previously been considered in small exoplanet atmospheres due to its potential as a biosignature photosynthetic product in reducing conditions, though it was deemed unlikely to occur \citep{Seager13Silane}. In the case studied in this work, silane would instead be a product of magma-atmosphere interactions.

We assume throughout this work that the condensate, in our case liquid SiO$_2$, is present in small enough quantities that it does not affect the atmospheric properties, such as, e.g., its opacity or heat capacity. However, some liquid must remain suspended in the gas. Lofting of condensates could alter the atmospheric profile, due to the liquid's larger specific heat \citep{Graham21}. Such condensate retention would also by definition produce clouds, which we do not explicitly model but which may have dramatic effects on the observability of these features. Therefore, such lofting should be investigated further, despite its reliance on complex microphysical processes. Possibly helpful analogs include more massive bodies such as hot Jupiters and brown dwarfs, in which silicate clouds have been previously studied \citep[e.g.][]{Burningham21, GaoPowell2021}.

\subsection{Other species and compositions}\label{sec:otherspecies}
We considered a chemical network including three reactions, in order to demonstrate the potential importance of these reactions on the atmospheric structure. Chemical networks involving more species and reactions have been employed in the study of sub-Neptune atmospheres in general, as well as magma-atmosphere interactions in various planetary contexts \citep[e.g.][]{Moses2013, SY22, Zilinskas2023}, though these works did not couple the products of silicate-hydrogen reactions with atmospheric structure in the sub-Neptune regime as we have here. In this section, we highlight possible extensions of the chemistry we consider that could influence the structure and evolution of sub-Neptune planets.

First, we considered an interior composed of pure SiO$_2$. However, including a more realistic, Earth-like composition would alter the products outgassed into the atmosphere. For example, if the rocky interior were approximated as pure MgSiO$_3$ instead, we would expect the evaporation products to be atomic Mg, SiO, and O$_2$, in equal molar proportions. This increase in the proportion of oxygen to silicon would likely lead to increased water production compared to our model. However, experimental results indicate that silicon may preferentially dissolve into H$_2$ compared to magnesium \citep{Shinozaki13, Shinozaki16}, potentially affecting, e.g., the Mg/Si ratio we would predict. We also do not consider that more complex chemistry in the melts will lower the activities of the relevant melt species and thus affect  the equilibrium vapor pressures of SiO and O$_2$, which may in turn alter the physics of convection inhibition. This example illustrates that interior composition can alter the atmospheric composition of sub-Neptunes, but that such interactions may be complex.

We do not expect an ice layer to form at the base of the atmosphere. Extrapolations indicate water and hydrogen are likely miscible at these temperatures and pressures, though experiments are needed to confirm this \citep{BaileyStevenson2021}. Additionally, solid phases of water are not stable at the temperatures and pressures at the base of the atmosphere \citep[e.g.][]{Madhusudhan20}. In fact, the temperatures and pressures are above the critical temperature and pressure of pure water, though the bulk behavior depends on the properties of the hydrogen-water-silane mixture, which is poorly constrained, not those of water alone \citep{Markham22}. The critical temperature of the silicate interior is likely higher than the 5000~K base temperature we assume \citep[e.g][]{XiaoStixrude2018}, indicating the overall mixture may not be super-critical, but more modeling of such mixtures is needed. If the mixture is super-critical, condensation of silicate vapor can no longer occur, and so convection would no longer be inhibited in the super-critical region of the envelope \citep{Markham22, Pierrehumbert2023}.

While not yet super-critical, the incompressibility of the melt and gasses at high pressures can lead to changes to the Gibbs free energies of formation not captured in our equilibrium model, which assumes ideal gasses and melt behavior. The equation of state of silicate melt is relatively well-constrained \citep[e.g.][]{deKokerStixrude2009}, allowing calculation of its molar volume as a function of pressure and therefore its change in chemical potential \citep[e.g.][]{SY22}. We find the change in chemical potential of the melt at the base of our atmosphere to be approximately 230 kJ/mol, enough to significantly alter the equilibrium state. However, this change in chemical potential may be compensated by non-ideal behavior in the product gaseous species on the other side of Reaction~\ref{eq:SiOrxn}. Specifically, an increase in the fugacity coefficient of SiO of approximately 40 to 100 at the conditions of the base of the atmosphere ($P \sim 10^5$~bar, $T \sim 5000$~K) would balance the $PV$ effects on the chemical potential of the melt. Unfortunately, the equations of state, and thus the fugacities, of the silicon vapor species we consider are not well-constrained. The observed behavior of other species, such as water vapor, indicate that such fugacities are plausible at the pressure conditions we consider \citep[e.g.][]{OtsukaKarato2011}. However, their precise determination is beyond the scope of this work.

We also ignore ingassing reactions, such as the solubility of hydrogen and water into the interior. Such reactions may be important in driving the interior composition and long-term atmospheric evolution of sub-Neptunes and super-Earths, although the relevant solubilities are highly uncertain \citep[e.g.][]{CS18, OlsonSharp2019, DornLichtenberg2021, SY22}.

Finally, we assume that the entire atmosphere, i.e. a given mass of hydrogen gas, has chemically equilibrated with the underlying magma ocean, similar to previous work on magma-hydrogen interactions \citep[e.g.][]{SY22, Markham22, Zilinskas2023}. This chemical processing leads to the sub-solar abundances we predict in the outer atmosphere: most of the oxygen and silicon is segregated to the deep atmosphere where it is more thermodynamically favorable. The result is an atmosphere that is overall strongly super-solar in e.g. oxygen abundance – O makes up 2 percent of the total atmosphere by number – but with strong variations with depth. The likelihood of reaching full chemical equilibrium depends on the mixing efficiency in the atmosphere during and after formation. If there is sufficient transport between atmospheric layers over the age of the planet, then we would expect chemical equilibrium to be achieved throughout the whole planet system, as modeled here. Conversely, higher elemental abundances of Si and O in the outer atmosphere than predicted here could indicate that chemical equilibrium between the outer and inner layers of the planet has not been reached, and therefore that the interior and atmosphere are poorly coupled chemically \citep[e.g.][]{Zilinskas2023}. Enhancement in these species could come from a higher metallicity in the initial accreted gas, although some of these enhanced metals would have condensed into refractory materials. Another source could be the ablation of accreting solid material; however, ablation is thought to occur in deeper regions than those probed by spectroscopic observations \citep[e.g.][]{BrouwersOrmel2020}.

\subsection{Time evolution}
Atmospheric loss processes are expected to affect a significant portion of sub-Neptunes, so the interplay between such processes and any evolution of the atmospheric composition should be considered carefully. We find that compositional gradients at depth tend to shrink the overall planet radius, if everything else is kept constant. A smaller radius tends to inhibit atmospheric loss, as the atmosphere must be removed from deeper within the gravitational potential well. 
However, atmospheric accretion dictates that planets' initial radiative-convective boundaries are close to the Bondi radius \citep[e.g.][]{GSS16} and deep non-convective layers slow thermal contraction over time \citep{MS22}. Therefore, whether the overall effect of the compositional gradients examined here inhibits or furthers atmospheric loss remains to be determined, highlighting the need for fully self-consistent accretion and loss models.

Another possibility is that hydrogen could be preferentially lost, increasing the molecular weight of the atmosphere \citep[e.g.][]{Malsky23}. The two major loss processes thought to shape exoplanet demographics, core-powered mass loss and photoevaporation, are hydrodynamic. Their strong winds are sufficient to drag along heavier species such as those we find could be present in this study \citep[e.g.][]{MS21}. However, due to the gradient in molecular weight, a wind escaping from the top of the atmosphere is relatively less enriched in outgassed species than the mean overall atmosphere, which could produce effective fractionation. Additionally, as hydrogen is depleted, the chemistry of the atmosphere could change, with the magma-hydrogen interaction producing more oxidized species \citep[e.g.][]{Zilinskas2023}. A model which simultaneously evolves the thermal state, atmospheric composition, and atmospheric mass could investigate these potential interplays further.

\section{Conclusions}
We analyze the chemical equilibrium between hydrogen gas and species vaporized from the surface of a silicate magma ocean, a condition likely to be present in the depths of young sub-Neptune planets. We find that SiO and O$_2$ react with hydrogen gas to produce significant amounts of silane (SiH$_4$) and water vapor (H$_2$O). The resulting depletion of SiO$_2$ draws more magma ocean products into the atmosphere, greatly increasing the atmosphere's silicon content compared to a model which does not account for the reducing conditions of the atmosphere. The amounts and proportions of these products likely vary depending on the chemical state of the interior. This implies that the atmospheric compositions of planets with magma oceans are a window into their interior composition, a promising prospect since such atmospheric abundances are observable with current and future telescopes. The chemical products of magma-atmosphere interaction in turn alter the atmospheric structure, inhibiting convection in the interior to a larger extent than previously found. These results imply that the presence of a magma ocean must be considered in order to understand the chemical abundances and overall atmospheric mass fractions of sub-Neptune planets.

\section*{Acknowledgements}
We thank James Rogers, Namrah Habib, and the anonymous reviewer for useful comments which improved the manuscript.
In this work we use the \textsc{numpy} \citep{numpy}, \textsc{matplotlib} \citep{Matplotlib}, and \textsc{scipy} \citep{scipy} packages. 
This research has been supported in part by the Alfred P. Sloan Foundation under grant G202114194 as part of the AEThER collaboration and by NASA under grant number 80NSSC21K0392 issued through the Exoplanet Research Program.

%%%%%%%%%%%%%%%%%%%%%%%%%%%%%%%%%%%%%%%%%%%%%%%%%%
\section*{Data Availability}

Data available upon request.

%%%%%%%%%%%%%%%%%%%% REFERENCES %%%%%%%%%%%%%%%%%%

% The best way to enter references is to use BibTeX:

\bibliographystyle{mnras}
\bibliography{si_vapor_2} 

%%%%%%%%%%%%%%%%%%%%%%%%%%%%%%%%%%%%%%%%%%%%%%%%%%

%%%%%%%%%%%%%%%%% APPENDICES %%%%%%%%%%%%%%%%%%%%%

\appendix
\section{Equilibrium constant values}\label{sec:Keq_plots}
\begin{figure}
	\includegraphics[width=\columnwidth]{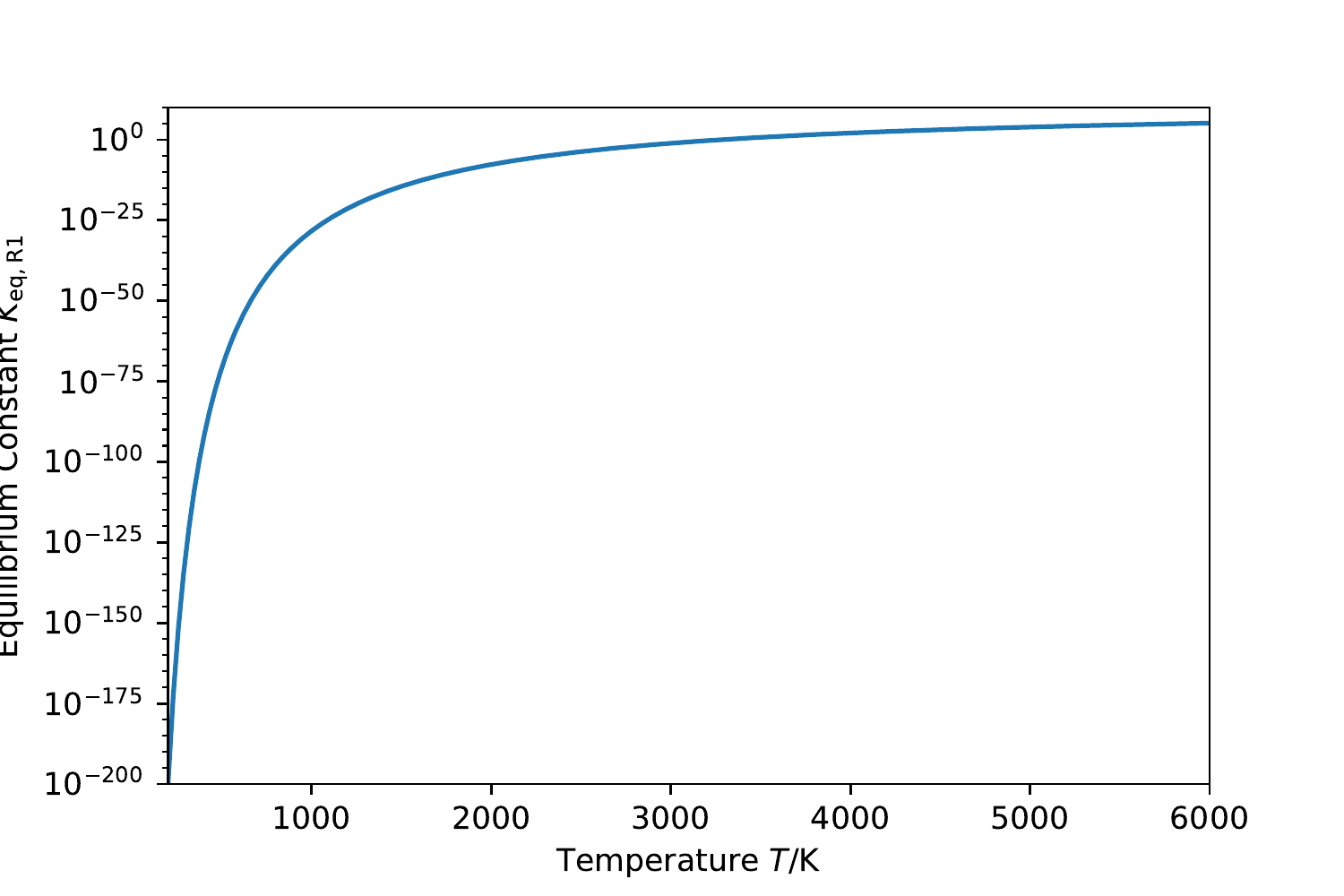}
    \caption{The equilibrium constant describing reaction~\ref{eq:SiOrxn}, as defined in Eq.~\ref{eq:SiOK}, as a function of temperature.}
    \label{fig:K_eqR1}
\end{figure}

\begin{figure}
	\includegraphics[width=\columnwidth]{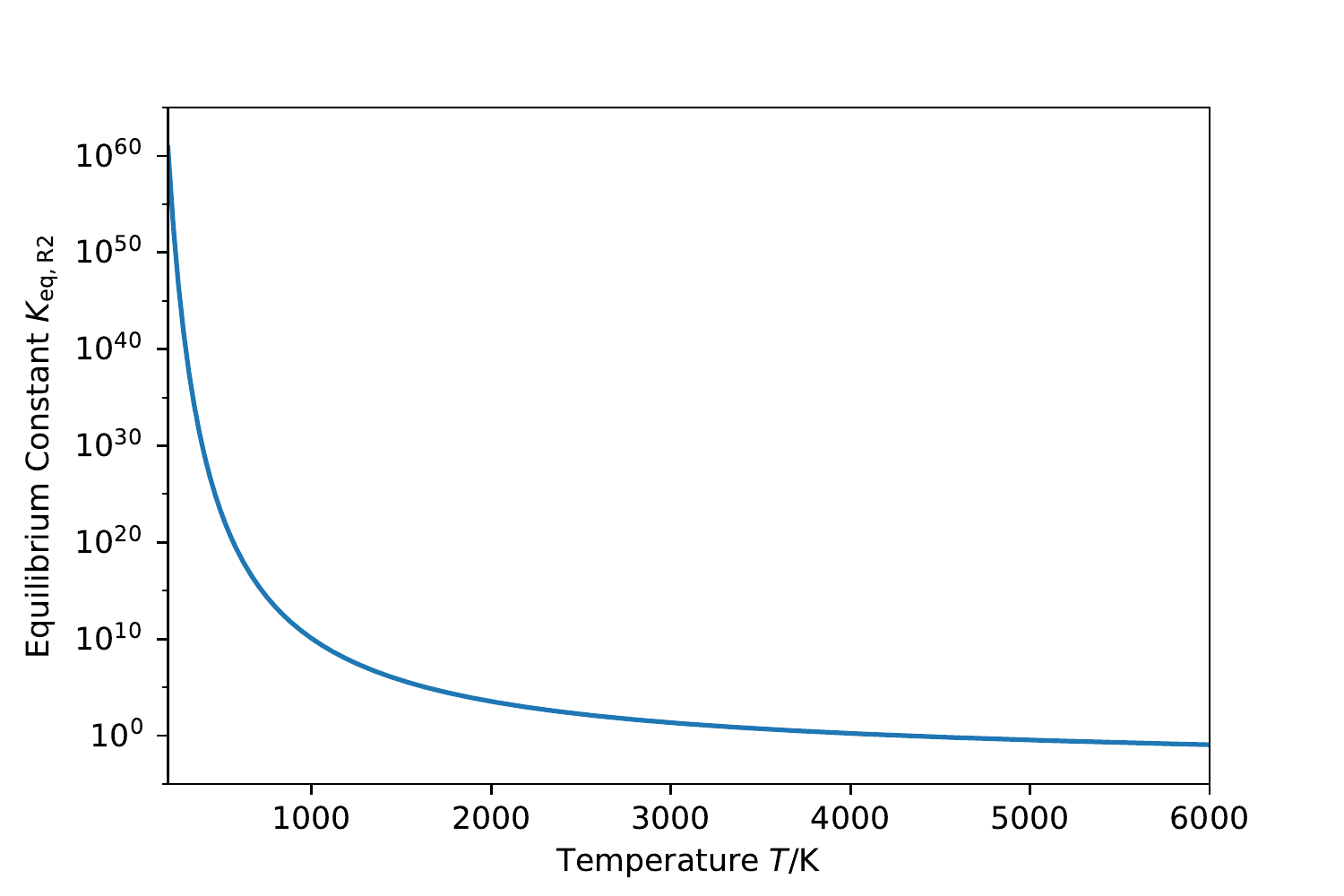}
    \caption{The equilibrium constant describing reaction~\ref{eq:O2rxn}, as defined in Eq.~\ref{eq:O2K}, as a function of temperature.}
    \label{fig:K_eqR2}
\end{figure}

\begin{figure}
	\includegraphics[width=\columnwidth]{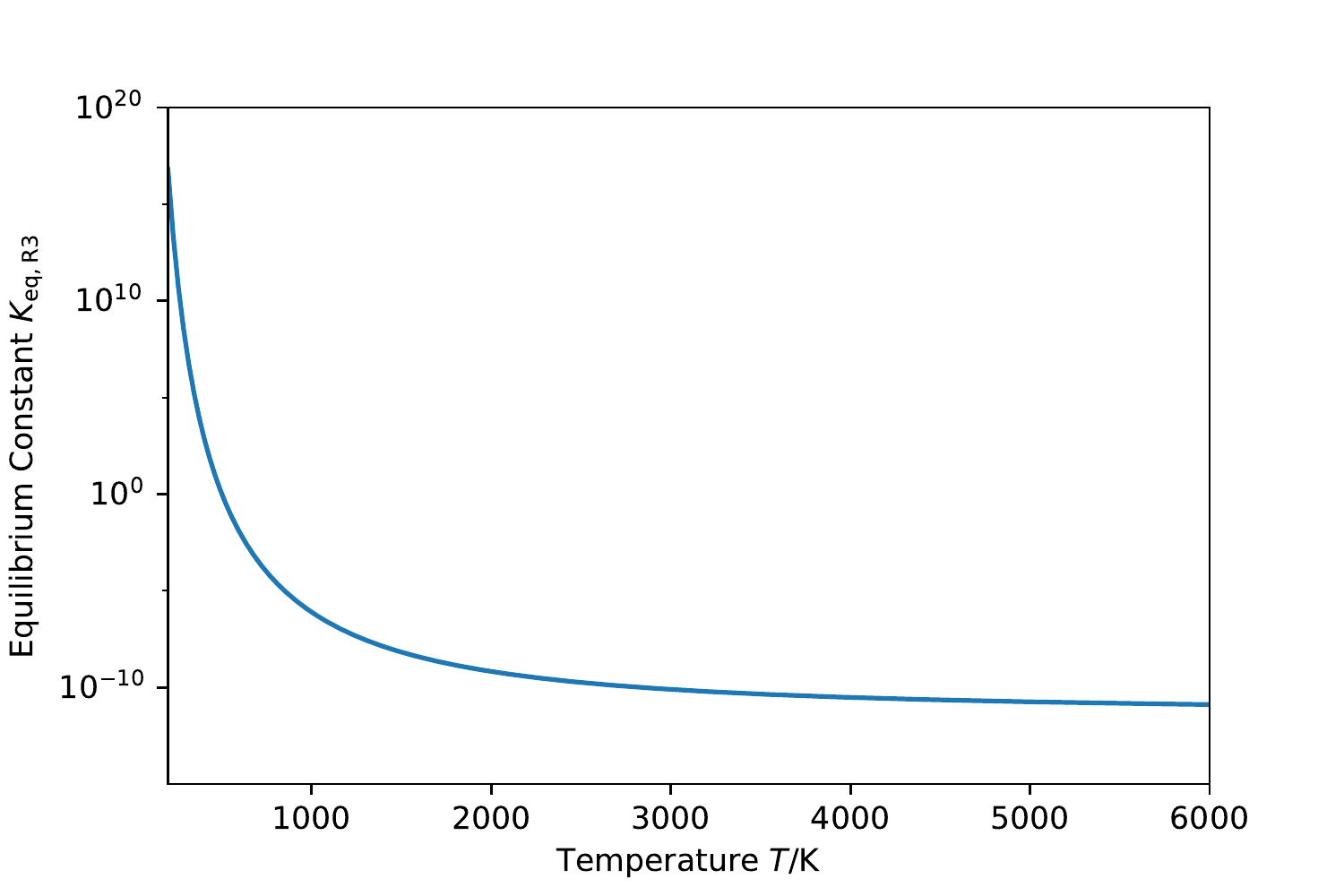}
    \caption{The equilibrium constant describing reaction~\ref{eq:SiH4rxn}, as defined in Eq.~\ref{eq:SiH4K}, as a function of temperature.}
    \label{fig:K_eqR3}
\end{figure}

In Figures~\ref{fig:K_eqR1}, ~\ref{fig:K_eqR2}, and ~\ref{fig:K_eqR3}, we plot the equilibrium constants we use in this work as functions of temperature. As described in Section \ref{sec:chem}, all values are calculated from the Gibbs free energies in NIST.

\section{Derivation of partial pressures}\label{sec:partialpress}
% See notes from 9/16/22 for derivation
As stated in Section~\ref{sec:chem}, Eqs.~\ref{eq:SiOK}, \ref{eq:O2K}, \ref{eq:SiH4K}, and \ref{eq:Nrelation}, along with the total pressure $P$ and temperature $T$, are sufficient to solve for the partial pressures of all components of the atmosphere at a given level, specifically $P_\mathrm{H_2}$, $P_\mathrm{H_2O}$, $P_\mathrm{SiO}$, $P_\mathrm{SiH_4}$, and $P_\mathrm{O_2}$. In practice, we use the following method, though many alternative derivations are possible in principle.

We begin by assuming an oxygen partial pressure, $P_\mathrm{O_2}$. We choose to guess this value because it varies by many orders of magnitude over a typical atmospheric profile, the most of any of the unknowns in our problem. Therefore, small changes in the other inferred partial pressures lead to large changes in the inferred $P_\mathrm{O_2}$, making it more amenable to solve for via numerical techniques than the other values, which vary slowly and therefore lead to failures to converge. Given the assumed $P_\mathrm{O_2}$ and the temperature, which yields values for all three equilibrium constants, it is easy to use Eq.~\ref{eq:SiOK} to solve for the partial pressure of SiO:
\begin{equation}
    P_\mathrm{SiO} = K_\mathrm{eq, R1}/P_\mathrm{O_2}^{1/2}.
\end{equation}

Similarly, we can invert Eq.~\ref{eq:O2K} to solve for the partial pressure of water as a function of the partial pressure of H$_2$, which is as yet unknown:
\begin{equation}\label{eq:PH2O}
    P_\mathrm{H_2O} = K_\mathrm{eq, R2} P_\mathrm{H_2} P_\mathrm{O_2}^{1/2}.
\end{equation}
Inserting Eq.~\ref{eq:PH2O} into Eq.~\ref{eq:SiH4K}, we can solve for $P_\mathrm{SiH_4}$:
\begin{equation}\label{eq:PSiH4wH}
    P_\mathrm{SiH_4} = \frac{K_\mathrm{eq, R3} P_\mathrm{SiO} P_\mathrm{H_2}^2}{K_\mathrm{eq, R2} P_\mathrm{O_2}^{1/2}},
\end{equation}
which also depends on the unknown $P_\mathrm{H_2}$. We can leverage the definition of $P$ as the sum of all partial pressures to express $P_\mathrm{H_2}$ in terms of the other species:
\begin{equation}
    P_\mathrm{H_2} = P - P_\mathrm{H_2O} - P_\mathrm{SiO} - P_\mathrm{O_2} - P_\mathrm{SiH_4}.
\end{equation}
Inserting Eq.~\ref{eq:PH2O} into this equation and collecting terms of $P_\mathrm{H_2}$ allows us to solve for $P_\mathrm{H_2}$ as a function of known values and $P_\mathrm{SiH_4}$:
\begin{equation}\label{eq:PH2}
    P_\mathrm{H_2} = \frac{P - P_\mathrm{SiO} - P_\mathrm{O_2} - P_\mathrm{SiH_4}}{1 + K_\mathrm{eq, R2} P_\mathrm{O_2}^{1/2}}.
\end{equation}
Such an expression allows us to substitute $P_\mathrm{H_2}$ in Eq.~\ref{eq:PSiH4wH}, eliminating all other unknown variables. This yields a simple quadratic equation:
\begin{equation}\label{eq:PSiH4quad}
    P_\mathrm{SiH_4} = \alpha (\beta - P_\mathrm{SiH_4})^2
\end{equation}
where 
\begin{equation}
    \alpha \equiv \frac{K_\mathrm{eq, R3} P_\mathrm{SiO}}{K_\mathrm{eq, R2} P_\mathrm{O_2}^{1/2} \big(1 + K_\mathrm{eq, R2} P_\mathrm{O_2}^{1/2}\big)^2}
\end{equation}
and
\begin{equation}
    \beta \equiv P - P_\mathrm{SiO} - P_\mathrm{O_2}
\end{equation}
are functions of known quantities defined for simplicity. Eq.~\ref{eq:PSiH4quad} can be solved using the quadratic formula:
\begin{equation}
    P_\mathrm{SiH_4} = \frac{2 \alpha \beta + 1 \pm \sqrt{4 \alpha \beta +1}}{2 \alpha}.
\end{equation}
Mathematically, this yields two solutions; however, in all cases only the negative branch is physically reasonable (i.e., yields $P_i > 0$ for all species). Once $P_\mathrm{SiH_4}$ is known, $P_\mathrm{H_2}$ can be solved via Eq.~\ref{eq:PH2}, and $P_\mathrm{H_2O}$ can be solved via Eq.~\ref{eq:PH2O}. This yields a full set of partial pressures, which correctly sum to the total pressure, and conform to chemical equilibrium at a given temperature, for a guessed $P_\mathrm{O_2}$. The only constraint not yet used is the number of atoms constraint, Eq.~\ref{eq:Nrelation}. We now calculate the ``atomic partial pressures'' using Eqs.~\ref{eq:sumPSi} and \ref{eq:sumPO}, and calculate the difference $\sum P_\mathrm{O} - \sum P_\mathrm{Si}$. This difference will vary as a function of the input $P_\mathrm{O_2}$, with one unique solution where the difference is zero. We solve for this $P_\mathrm{O_2}$ which balances the reactions correctly numerically for each layer, using the \texttt{fsolve} function of the \textsc{scipy.optimize} package \citep{scipy}. In practice, since we solve the atmospheric structure layer-by-layer in small steps, the previous value of $P_\mathrm{O_2}$ provides a good starting guess for the value in the next layer deeper, which we use to increase computational speed compared to a fully naive guess.

\section{Derivation of multi-species convection criterion}\label{sec:criterion}
% See notes from 3/11/22
As discussed in Section~\ref{sec:structure}, in this work we derive a multi-species convection criterion, which turns out to be the sum of the individual convection criteria of each species, $\sum_i c_i \geq 1$, where $c_i$ is given by Eq.~\ref{eq:conv_crit}. It is non-trivial that the overall convection criterion is the sum of those of each species, so we demonstrate that here. For demonstration purposes, we consider the case of two condensables, but the argument applies to $n$ condensables equally well. Let us define a three component atmosphere, composed of a dry component with mass mixing ratio $q_\mathrm{d}$ and molecular weight $\mu_\mathrm{d}$, and two condensable species with mass mixing ratios $q_1$ and $q_2$ and molecular weights $\mu_1$ and $\mu_2$ respectively. By definition, $q_\mathrm{d} + q_1 + q_2 = 1$. Meanwhile, the overall mean molecular weight $\mu$ is given by
\begin{equation}\label{eq:mu_d12}
    \mu = \frac{\mu_\mathrm{d} \mu_1 \mu_2}{\mu_1 \mu_2 +  (\mu_\mathrm{d} \mu_2 - \mu_1 \mu_2) q_1 + (\mu_\mathrm{d} \mu_1 - \mu_1 \mu_2) q_2}.
\end{equation}

For convection to be inhibited, the density gradient in the environment must be steeper than that of a parcel moved adiabatically \citep[e.g.][]{L17}:
\begin{equation}\label{eq:inhib}
    \bigg(\pdv{\ln T}{\ln P} - \pdv{\ln \mu}{\ln P}\bigg)_\mathrm{env} > \bigg(\pdv{\ln T}{\ln P} - \pdv{\ln \mu}{\ln P}\bigg)_\mathrm{ad}.
\end{equation}
Therefore, to assess whether convection operates in a regime with multiple species changing in abundance, we must compute the change in molecular weight with pressure $\pdv*{\ln \mu}{\ln P}$. The molecular weight changes with pressure for two reasons: due to changes in the abundance of species 1, or due to changes in the abundance of species 2. Quantitatively, we can express this as a sum of partial derivatives with the mixing ratio of the other species held constant:

\begin{equation}\label{eq:dmudP_sum}
    \pdv{\ln \mu}{\ln P} = \pdv{\ln \mu}{\ln P}\bigg|_{q_2}(q_1) + \pdv{\ln \mu}{\ln P}\bigg|_{q_1}(q_2).
\end{equation}
Here the first term is the change in molecular weight due to changing $q_1$, with $q_2$ held fixed, and the second term is the change in molecular weight due to changing $q_2$, with $q_1$ held fixed.

Examining the first term in Eq.~\ref{eq:dmudP_sum}, it can be expanded into the product of two partial derivatives
\begin{equation}\label{eq:dmudP}
    \pdv{\ln \mu}{\ln P}\bigg|_{q_2} = \pdv{\ln \mu}{\ln q_1}\bigg|_{q_2} \pdv{\ln q_1}{\ln P}\bigg|_{q_2}.
\end{equation}
The first term in the product can be computed using the definition of $\mu$ in Eq.~\ref{eq:mu_d12} and simplifies to
\begin{equation}\label{eq:dmudq}
    \pdv{\ln \mu}{\ln q_1}\bigg|_{q_2} = \mu q_1 \bigg(\frac{1}{\mu_\mathrm{d}} - \frac{1}{\mu_1} \bigg).
\end{equation}
Meanwhile, the second term in Eq.~\ref{eq:dmudP} can be expressed as the sum of two terms: the change in condensable mass mixing ratio as the total pressure changes at fixed temperature, and the change in condensable mass mixing ratio as the temperature changes due to the $T$--$P$ relation:
\begin{equation}\label{eq:dqdP}
    \pdv{\ln q_1}{\ln P}\bigg|_{q_2} = \pdv{\ln q_1}{\ln P}\bigg|_{T, q_2} + \pdv{\ln q_1}{\ln T}\bigg|_{P, q_2} \pdv{\ln T}{\ln P}
\end{equation}
The first term in this equation can be solved using the definition of the mass mixing ratio, $q_1 = \mu_1 P_1/(\mu P)$, and reduces to
\begin{equation}\label{eq:dqdP_T}
    \pdv{\ln q_1}{\ln P}\bigg|_{T, q_2} = - \frac{\mu_\mathrm{d}}{\mu}.
\end{equation}
We note that this term is negative, because increasing the total pressure without changing the temperature leaves the partial pressure of the condensable unchanged, therefore decreasing the mass mixing ratio.

The derivative of $q_1$ with respect to $T$, which appears in the second term of Eq.~\ref{eq:dqdP}, can also be solved by inserting the definition of the mass mixing ratio, but it simplifies considerably less and introduces cross-terms dependent on $q_2$:
\begin{equation}\label{eq:dqdT}
    \pdv{\ln q_1}{\ln T}\bigg|_{P, q_2} = \bigg[\frac{\mu_\mathrm{d}}{\mu} + q_2 \bigg(1 - \frac{\mu_\mathrm{d}}{\mu_2}\bigg)\bigg] \pdv{\ln P_1}{\ln T} - q_2 \bigg(1 - \frac{\mu_\mathrm{d}}{\mu_2}\bigg) \pdv{\ln P_2}{\ln T}.
\end{equation}
Eqs.~\ref{eq:dqdP_T} and \ref{eq:dqdT} can then be inserted into Eq.~\ref{eq:dqdP}, which can be substituted along with Eq.~\ref{eq:dmudq} into Eq.~\ref{eq:dmudP}. After some algebraic manipulation, this substitution leads to an expression for the overall gradient as a function of $q_1$ with $q_2$ fixed:
\begin{equation}
\begin{split}\label{eq:dmudPfull}
    \pdv{\ln \mu}{\ln P}\bigg|_{q_2} &= q_1 \bigg(1 - \frac{\mu_\mathrm{d}}{\mu_1}\bigg) \bigg[\pdv{\ln P_1}{\ln T}\pdv{\ln T}{\ln P}-1\bigg] \\
    &+ \frac{\mu}{\mu_\mathrm{d}} q_1 q_2 (1 - \frac{\mu_\mathrm{d}}{\mu_1}) (1 - \frac{\mu_\mathrm{d}}{\mu_2}) \pdv{\ln T}{\ln P} (\pdv{\ln P_1}{\ln T} - \pdv{\ln P_2}{\ln T}).
\end{split}
\end{equation}
As $q_1$ and $q_2$ are completely symmetrical in these equations, the expression for $\pdv*{\ln \mu}{\ln P}|_{q_1}$ can be obtained by swapping the 1 and 2 subscripts in Eq.~\ref{eq:dmudPfull}. Upon adding these equations together per Eq.~\ref{eq:dmudP_sum}, it is immediately clear that the second terms cancel, leaving
\begin{equation}\label{eq:dmudP_total}
\begin{split}
    \pdv{\ln \mu}{\ln P} &= q_1 \bigg(1 - \frac{\mu_\mathrm{d}}{\mu_1}\bigg)\bigg[\pdv{\ln P_1}{\ln T}\pdv{\ln T}{\ln P}-1\bigg] \\
    &+ q_2 \bigg(1 - \frac{\mu_\mathrm{d}}{\mu_2} \bigg) \bigg[\pdv{\ln P_2}{\ln T} \pdv{\ln T}{\ln P} - 1\bigg] \\
    &= \alpha_1 \gamma_1 \pdv{\ln T}{\ln P} - \alpha_1 (1 - \varpi_1 q_1 - \varpi_2 q_2) \\
    &+ \alpha_2 \gamma_2 \pdv{\ln T}{\ln P} - \alpha_2 (1 - \varpi_1 q_1 - \varpi_2 q_2)
\end{split}
\end{equation}
using the notation of \citet{L17}, where
\begin{equation}
    \alpha_i \equiv \mu q_i \bigg(\frac{1}{\mu_\mathrm{d}} - \frac{1}{\mu_i}\bigg),
\end{equation}
\begin{equation}
    \gamma_i \equiv \frac{\mu_\mathrm{d}}{\mu}\pdv{\ln P_i}{\ln T},
\end{equation}
and
\begin{equation}
    \varpi_i \equiv 1 - \frac{\mu_\mathrm{d}}{\mu_i}.
\end{equation}

Finally, we can insert Eq.~\ref{eq:dmudP_total} into Eq.~\ref{eq:inhib} to obtain the convection criterion. We note that the second and fourth terms of Eq.~\ref{eq:dmudP_total} will be the same on either side of the inequality in Eq.~\ref{eq:inhib} and therefore cancel, eliminating all further cross terms. This leaves the inequality as
\begin{equation}
    \bigg[\bigg(\pdv{\ln T}{\ln P}\bigg)_\mathrm{env} - \bigg(\pdv{\ln T}{\ln P}\bigg)_\mathrm{ad}\bigg] (1 - \alpha_1 \gamma_1 - \alpha_2 \gamma_2) > 0.
\end{equation}
Since the environmental temperature gradient will not be less than the adiabatic gradient, this inequality implies that
\begin{equation}
    1 > \alpha_1 \gamma_1 + \alpha_2 \gamma_2
\end{equation}
where $\alpha_i \gamma_i = \chi_i$ as defined in Eq.~\ref{eq:conv_crit}: the inhibition criteria solely add.

%%%%%%%%%%%%%%%%%%%%%%%%%%%%%%%%%%%%%%%%%%%%%%%%%%

% Don't change these lines
\bsp	% typesetting comment
\label{lastpage}
\end{document}